\documentclass{mcom-l}

\usepackage{bm}
\usepackage{mathrsfs}
\usepackage{amssymb}
\usepackage{enumitem}
\usepackage{times}
\usepackage{pdfpages}
\usepackage{float}
\usepackage{subfig,graphicx}
\usepackage{csquotes}
\usepackage{hyperref}
\usepackage{amsmath}
\usepackage{ulem}
\usepackage{lipsum} 
\usepackage{nccmath}
\usepackage{bigints}
\usepackage{natbib}
\usepackage{multirow}
\usepackage{adjustbox}
\usepackage{threeparttable}
\usepackage{booktabs}
\usepackage{bbm}

\usepackage{pdfpages}
\usepackage{wrapfig}
\usepackage{color}

\usepackage{float}
\usepackage{tabularx}
\usepackage{caption}
\usepackage{cite}

\usepackage{graphicx}

\newtheorem{theorem}{Theorem}
\newtheorem{lemma}{Lemma}
\newtheorem{corollary}{Corollary}

\newtheorem{remark}{Remark}

\numberwithin{equation}{section}
\numberwithin{theorem}{section}
\numberwithin{corollary}{section}
\numberwithin{lemma}{section}
\numberwithin{proposition}{subsection}
\numberwithin{corollary}{section}
\numberwithin{example}{section}
\numberwithin{remark}{section}

\begin{document}

\title[Asymptotic Theory of Tail Dependence and Bootstrap for Checkerboard Copulas]{Asymptotic Theory of Tail Dependence and Bootstrap for Checkerboard Copulas}

\author{Mayukh Choudhury}
\address{Department of Mathematics, Indian Institute of Technology Bombay, Mumbai 400076, India}
\email{214090002@iitb.ac.in}

\author{Debraj Das}
\address{Department of Mathematics, Indian Institute of Technology Bombay, Mumbai 400076, India}
\email{debrajdas@math.iitb.ac.in}
\author{Sujit Ghosh}
\address{Departement of Statistics,
  North Carolina State University, Raleigh, NC 27695, USA}
\email{sujit.ghosh@ncsu.edu}





\begin{abstract}
A comprehensive asymptotic and bootstrap theory is established for checkerboard-based estimation of the copula and its lower and upper tail copula counterparts under unknown marginal distributions. The proposed estimator of the tail copula extends a local bilinear interpolation of the empirical copula to the tail region, providing a flexible nonparametric approach for modeling extremal dependence. Almost sure uniform consistency is established under mild conditions on the checkerboard grid. Weak convergence of the checkerboard copula process is derived, showing that smoothing preserves the first-order asymptotic limit of the empirical copula process, including the effect of marginal estimation. These results are further extended to lower and upper tail copula processes, yielding asymptotic normality for tail dependence measures. Since the limiting processes depend on unknown characteristics of the underlying true copula, a multiplier bootstrap procedure adapted to the checkerboard structure is proposed and shown to be asymptotically valid. Simulation studies and statistical applications validates our theoretical findings 
under a range of dependence structures. Although the limiting processes match with that observed for the empirical copula, the finite sample performance shows a noticeable improvement under checkerboard smoothing.
\end{abstract}

\maketitle

\section{Introduction}\label{sec:intro}
Copulas have emerged as fundamental tools for modeling dependence in multivariate data.  
Unlike classical correlation measures that capture only linear association, copulas provide a way to describe the entire dependence structure between random variables independent of their marginal distributions. This flexibility has led to widespread use of copulas in a variety of fields, including finance, insurance, hydrology, and environmental science, where understanding joint tail behavior is of paramount importance (cf. \citet{KarolyiStulz1996,LonginSolnik2001,CampbellKoedijkKofman2002,CaillaultGuegan2005} and references there in). A cornerstone of the theory is Sklar’s theorem (cf.  \citet{Sklar1959}), which states that any multivariate distribution function \(H\) with marginals \(F\) and \(G\) can be expressed as
\[
H(x,y) = C\!\big(F(x),\,G(y)\big),
\]
where \(C:[0,1]^2\to[0,1]\) is a copula function. If \(F\) and \(G\) are continuous, then \(C\) is unique. This result allows one to separate marginal behavior from dependence structure: one models the marginals and the copula independently, and then combines them to understand joint behavior. In finance, for example, copula models have been used to capture dependence in asset returns and credit portfolio losses (cf.\citet{CherubiniEtAl2004, Joe2014}). In hydrology, copulas model joint extremes of rainfall and river discharge (cf. \citep{SalvadoriDeMichele2004}). These applications underscore that accurate estimation of copulas and their tail behavior is critical in risk assessment and multivariate analysis.

Given the prominence of copulas, their nonparametric estimation has been extensively studied. The classical empirical copula, was originally formalized by \citet{Deheuvels1979, Deheuvels1981}. 
The empirical copula denoted by $\hat C_n(u,v)$ is the proportion of marginal ranks of the observed data that fall in the rectangle $[0,u]\times[0,v]$ and serves as a nonparametric estimator of the underlying copula $C(u,v)$. The empirical copula is a natural nonparametric estimator of the true copula \(C\). Under appropriate regularity conditions, it is uniformly consistent and admits a functional central limit theorem (cf. \citet{fermanian2004weak, Segers2012}), providing a basis for inference on dependence structure. Despite its theoretical appeal, the empirical copula has several limitations in practice. It is discrete by construction and may exhibit erratic behavior, particularly near the boundaries of the unit square. This can be problematic when estimating smooth functionals of the copula, such as tail dependence measures or when constructing confidence bands. In addition, the empirical copula sometimes suffers from edge effects and boundary discontinuities that complicate functional limit theory and bootstrap approximations (cf. \citet{Segers2012,BerghausVolgushev2017,SegersSibuyaTsukahara2017}).

 To circumvent the shortcomings of the empirical copula, a variety of smoothed copula estimators have been proposed in the literature. Early approaches include kernel–based smoothing of the empirical copula process \citet{GijbelsMielniczuk1990, GeenensCharpentierPaindaveine2017}, Bernstein polynomial based copula estimators \citet{SancettaSatchell2004, janssen2012bernstein}. Other methodologies, such as wavelet or orthogonal polynomial–based estimators, have also been considered in recent work of \citet{ProvostZang2024}. These estimators aim to improve finite–sample performance and to produce smoother estimates that facilitate functional inference and bootstrap procedures. However, kernel–based approaches often require careful bandwidth selection and can suffer from boundary bias, especially near the edges of the unit square (cf. \citet{WenWu2018,MuiaAtuteyHasan2025}). Bernstein and other polynomial methods, while smooth, can be computationally demanding in higher dimensions and may introduce unintended bias in tail regions when the degree of approximation is not carefully chosen (cf. \citet{SancettaSatchell2004}). 

 An alternative class of smooth copula estimators is based on grid–based interpolation of the empirical copula. Among these, checkerboard copulas have attracted considerable attention because of their simplicity, continuity, and minimal tuning requirements. The method partitions $[0, 1]^2$ into a grid (we denote the grid size by $m$) and constructs a continuous copula through bilinear interpolation of empirical copula values on the grid points; see Section \ref{sec:checkerboard} for details. Unlike kernel methods, checkerboard estimators avoid bandwidth selection while preserving key copula properties.
 
Early contributions include \citet{CuberosMasielloMaume2020}, who employed checkerboard approximations in quantile estimation for sums of dependent random variables, and \citet{GonzalezBarriosHoyos2021}, who established error bounds and convergence results for sample d-copula approximations. Subsequent work expanded both the theoretical and applied scope of checkerboard copulas. \citet{GriessenbergerJunkerTrutschnig2022} and \citet{JunkerGriessenbergerTrutschnig2021} developed dependence measures and showed that checkerboard estimators provide consistent nonparametric estimation of multivariate and asymmetric dependence structures. Weak convergence results for multilinear empirical copula processes were studied by \citet{GenestNeslehovaRemillard2014}, while \citet{LinWangZhangZhao2025} justified checkerboard copulas through an entropy-maximization perspective. More recently, \citet{LuGhosh2024} proposed a smoothed checkerboard Bernstein–sieve estimator for conditional copulas. These developments, together with applications in empirical process theory and multiplier bootstrap methods for tail copulas, motivate the checkerboard-based framework adopted in this paper.

These developments suggest that checkerboard-type smoothing can improve both finite-sample performance and theoretical tractability. However, most existing work has focused on global copula estimation, with comparatively limited attention to asymptotic inference for tail copulas under checkerboard smoothing. Tail copulas and stable tail dependence functions play a central role in extreme value theory and risk management because they characterize extremal dependence and joint tail behavior. Foundational work by \citet{Huang1992} introduced nonparametric estimation of stable tail dependence functions, while \citet{DreesHuang1998} established convergence rates and optimality properties. Connections between stable tail dependence functions and tail copulas, together with weak convergence results for empirical tail copula processes, were developed by \citet{schmidt2006}. A comprehensive treatment of multivariate extremes and tail dependence is provided in \citet{deHaanFerreira2006}.

Subsequent research considered statistical inference and smoothing methods for tail dependence estimation. \citet{PengQi2007} and \citet{Peng2008Bootstrap} studied smooth estimation and bootstrap inference for stable tail dependence functions. Method-of-moments estimation was investigated by \citet{EinmahlKrajinaSegers2008}, while \citet{BucherDette2013} established weak convergence results for empirical tail copula processes under broad conditions. More recently, \citet{Kiriliouk2018} proposed empirical beta copula smoothing methods with improved finite-sample behavior, and \citet{EinmahlSegers2021} extended tail dependence inference to functional data settings. Despite these advances, the asymptotic behavior of tail copula estimators under checkerboard smoothing remains largely unexplored. These works provide the theoretical foundation for the checkerboard-based tail inference framework developed in this paper.

In Section \ref{sec:main} of this paper, we establish a comprehensive asymptotic theory for checkerboard–based tail copula estimation. Specifically, our contributions from Section \ref{sec:main} are as follows: 
\begin{enumerate}[label =(\Alph*)]
    \item We define a checkerboard–smoothed estimator $\hat{C}_n^{(m)}$ of the true copula $C$ and establish the almost sure uniform consistency of it in Theorem \ref{thm:consistency_checkerboard} which is
    \[
    \mathbf{P}\Big[\sup_{(u,v)\in[0,1]^2}
\bigl|
\widehat{C}_n^{(m)}(u,v) - C(u,v)
\bigr|
=
o(1)\Big] = 1,
\quad \text{as}\; n\to\infty.
    \]
\item We next derive that, under suitable smoothness
conditions, the checkerboard smoothing does not change asymptotic distribution. In particular, the empirical checkerboard copula
admits the same weak limit as the classical empirical copula process (cf. Theorem \ref{thm:weakconv_checkerboard}), i.e., 
\[
\sqrt{n}\bigl(\widehat{C}_n^{(m)} - C\bigr)
\;\rightsquigarrow\;
\mathbb{G}_C
\quad\text{in}\quad
\ell^\infty([0,1]^2),
\qquad\; \text{as}\; n\to\infty,
\]
where \(\mathbb{G}_C\) is the centered Gaussian process associated with the classical empirical copula and `$\rightsquigarrow$' denotes weak convergence. The space $\ell^\infty([0,1]^2)$ is the space of all bounded functions defined on $[0, 1]^2$.


\item Based on the empirical checkerboard copula $\hat{C}_n^{(m)}$, we introduce non-parametric checkerboard-based lower and upper tail copula estimators $\hat \Lambda_{L,n}^{(m)}$ and $\hat \Lambda_{U,n}^{(m)}$ of the true lower and upper tail copulas $\Lambda_L$ and $\Lambda_U$ (see section \ref{sec:checkerboard} for details). In Theorem \ref{thm:strongconstail}, we derive the strong consistency of these estimators which are given by
\[
\mathbf{P}\Big[\lim_{n\to\infty}d(\hat \Lambda_{L,n}^{(m)},\Lambda_L)=0\Big]=1  
\quad\text{and}\quad \mathbf{P}\Big[\lim_{n\to\infty}d(\hat \Lambda_{U,n}^{(m)},\Lambda_U)=0\Big]=1.
\]
Here all the tail copulas are considered as elements of the space $B_\infty(\overline{\mathbb{R}}_+^2)$ which consists of all functions $f: \overline{\mathbb{R}}_+^2 \to \mathbb{R}$ that are locally uniformly bounded on every compact subset of $\overline{\mathbb{R}}_+^2$. Throughout the paper, we denote that
\[
\overline{\mathbb{R}}_+^2
:=
[0,\infty]^2 \setminus \{(\infty,\infty)\}
=
\left\{(x,y)\in[0,\infty]^2 : (x,y)\neq(\infty,\infty)\right\}.
\] The metric $d(\cdot,\cdot)$ associated with the space $B_{\infty}(\bar{\mathbb{R}}_{+}^2)$ is defined as:
\[ d(f_1,f_2) = \sum_{i=1}^{\infty} 2^{-i} \big( \| f_1 - f_2 \|_{T_i} \wedge 1 \big),\]
where for $i \in \mathbb{N},$ we write, $T_{3i}= T_{3i-1} \cup [0,i]^2$, $T_{3i-1}= T_{3i-2} \cup ([0,i] \times \{\infty\})$, $T_{3i-2}= T_{3(i-1)} \cup (\{\infty\} \times [0,i])$,$T_0 = \emptyset$ and $\|f_1 - f_2\|_{T_i} = \sup_{(x,y) \in T_i} |f_1(x,y)-f_2(x,y)|$. Thus, a sequence $\{f_n\}_{n\ge 1} \in \mathcal B_\infty(\overline{\mathbb R}_+^2)$converges with respect to the metric $d$ if and only if it converges uniformly on each set $T_i$.

\item In Theorem \ref{thm:weakconv_tail_checkerboard}, we provide analogous weak convergence behavior for the upper and lower tail copula (see section \ref{sec:checkerboard} for definition), by demonstrating that, 
    \begin{align*}
&\sqrt{k_n}
\Big(
\widehat{\Lambda}_{L,n}^{(m)} - \Lambda_L
\Big)
\;\rightsquigarrow\;
\mathbb{G}_{\hat{\Lambda}_L},\quad\text{in}\; B_\infty(\overline{\mathbb{R}}_+^2),
\qquad\; \text{as}\; n\to\infty,\\
& \sqrt{k_n}
\Big(
\widehat{\Lambda}_{U,n}^{(m)} - \Lambda_U
\Big)
\;\rightsquigarrow\;
\mathbb{G}_{\hat{\Lambda}_U},
\quad\text{in } B_\infty(\overline{\mathbb{R}}_+^2),
\qquad \text{as}\; n\to\infty,
\end{align*}
where $\mathbb G_{\hat \Lambda_L}$, $\mathbb G_{\hat \Lambda_U}$ denote some centered Gaussian processes.
$k_n/n$ essentially denotes the proportion of sample used to derive $\widehat{\Lambda}_{L,n}^{(m)}$ and $\widehat{\Lambda}_{U,n}^{(m)}$. As an immediate consequence of this result, we derived the CLT for checkerboard-based tail copula coefficient in Corollary \ref{cor:asymp_normal_tailcoeff}, as:
\[
\sqrt{k_n}\,
\bigl(\hat \lambda_{L,n}^{(m)} - \lambda_L\bigr)
\;\overset{d}{\longrightarrow}\;
N\bigl(0, \sigma_L^2\bigr),
\qquad n\to\infty.
\] 
\end{enumerate}
The weak convergence results in Theorem~\ref{thm:weakconv_tail_checkerboard} establish a centered Gaussian limit for the empirical checkerboard tail copula process. However, the limiting covariance depends on unknown tail copulas and, in the unknown--marginal setting, on their partial derivatives, making direct asymptotic inference impractical. Bootstrap methods for copula processes have been widely studied; see, for example, \citet{fermanian2004weak,GenestRemillard2008,BucherDette2010,Segers2012} for empirical copulas and \citet{ChenFan2006} for smoothed copula estimators. Nevertheless, as noted by \citet{BucherDette2013}, standard bootstrap procedures are not directly suitable for tail copula inference because tail estimators depend only on the extreme observations with $k_n \ll n$. Consequently, resampling from the full empirical distribution fails to capture the correct tail variability, and the classical empirical bootstrap performs poorly for tail empirical processes; see also \citet{BucherDette2010}. To overcome these difficulties, \citet{BucherDette2013} proposed direct and partial-derivatives multiplier bootstrap methods for empirical tail copulas in the unknown--marginal case. Although both approaches are asymptotically valid, the partial-derivatives method requires additional smoothness assumptions and nonparametric estimation of derivatives in tail regions, which can be technically challenging and sensitive to tuning choices.

\medskip

In section \ref{sec:des} of this paper, we adopt and extend the direct multiplier bootstrap approach of \citet{BucherDette2013} to the empirical checkerboard copula as well as the empirical checkerboard tail copula. 
Multipliers are $n$ iid positive sub-exponential random variables $\xi_1, \dots, \xi_n$ with $\mathbf{E}(\xi_1)=\mu$ and $\text{var}(\xi_1)=\tau^2<\infty$. 
Multipliers in bootstrap is particularly
natural in our setting, since the checkerboard estimator already incorporates a smoothing mechanism via local convex combinations, and moreover the direct multiplier bootstrapping
avoids the need for explicit derivative estimation. By combining multiplier
reweighting with checkerboard smoothing, we develop a bootstrap procedure that
preserves the mean as well as the extremal structure of the data while accurately reproducing the
first--order stochastic fluctuations. 
Following are the contributions on bootstrap approximation of Section  \ref{sec:des}. 
\begin{enumerate}[label =(\Alph*),start=5]
\item First we construct $\hat C_n^{(m),\xi,\xi}$ as the bootstrap version of $\hat{C}_n^{(m)}$. We establish that conditionally on data \[
\frac{\mu}{\tau}\sqrt{n}\Big(\hat C_n^{(m),\xi,\xi}-\hat{C}_n^{(m)}\Big)\overset{\mathbf P}{\underset{\xi}{\rightsquigarrow}}\;\ 
\mathbb G_{C}
\quad
\text{in } \ell^\infty([0,1]^2),\quad \text{as}\; n \rightarrow \infty,
\]
where $\mathbb{G}_C$ is the centered tight Gaussian process, appeared as the limit of the original empirical checkerboard copula process. By the notation ``$\overset{\mathbf P}{\underset{\xi}{\rightsquigarrow}}$'' we mean, weak convergence conditioned on the data in probability.
\item Subsequently, we construct $\hat \Lambda_{L, n}^{(m),\xi,\xi}$ and $\hat \Lambda_{U, n}^{(m),\xi,\xi}$ as bootstrap versions of $\hat \Lambda_{L, n}^{(m)}$ and $\hat \Lambda_{U, n}^{(m)}$ respectively. We establish that conditionally on the data,
  \begin{align*}
&\frac{\mu}{\tau}\sqrt{k_n}\,
\Bigl(
\widehat{\Lambda}^{(m),\xi,\xi}_{L,n}
-
\widehat{\Lambda}^{(m)}_{L,n}
\Bigr)
\;\overset{\mathbf P}{\underset{\xi}{\rightsquigarrow}}\;
\mathbb{G}_{\hat{\Lambda}_L},\quad\text{in}\; B_\infty(\overline{\mathbb{R}}_+^2),
\qquad\; \text{as}\; n\to\infty,\\
& \frac{\mu}{\tau}\sqrt{k_n}\,
\Bigl(
\widehat{\Lambda}^{(m),\xi,\xi}_{U,n}
-
\widehat{\Lambda}^{(m)}_{L,n}
\Bigr)
\;\overset{\mathbf P}{\underset{\xi}{\rightsquigarrow}}\;
\mathbb{G}_{\hat{\Lambda}_U},
\quad\text{in } B_\infty(\overline{\mathbb{R}}_+^2),
\qquad \text{as}\; n\to\infty,
\end{align*}
where, the limits $\mathbb{G}_{\hat{\Lambda}_L}$ and $\mathbb{G}_{\hat{\Lambda}_U}$ are the centered tight Gaussian processes, mentioned in the weak convergence of the original empirical checkerboard tail copula processes. 
\end{enumerate}
The above two results provide a fully feasible bootstrap based inference procedure for the true copula and true tail copula processes utilizing the empirical checkerboard-based copula. The bootstrap procedure neither require explicit estimation of the covariance structure of the limiting Gaussian processes or require estimation of the derivatives of the underlying true copula. The practical usefulness of this bootstrap framework is illustrated in section~\ref{sec:applications} based on following two practical examples:
\begin{enumerate}[label =(\roman*)]
    \item testing of equality of two lower tail copulas coming out of two independent bivariate populations.
    \item testing goodness--of--fit of a parametric model to the underlying lower tail copula process.
\end{enumerate}
In either case, we show that our proposed bootstrap test is asymptotically of correct size and is consistent. We also study the finite sample performance of our proposed tests by comparing both size and power of our tests with that of the tests based on empirical copula proposed by \citet{BucherDette2013}. We observe that our tests based on empirical checkerboard copula outperforms the tests of \citet{BucherDette2013} in all the simulation settings. Therefore, the Checkerboard smoothing improves the reliability of size control as well as enhances the sensitivity to the tail dependence alternatives to the tests based on empirical copula.

\medskip

The rest of the paper is organized as follows. In section~\ref{sec:checkerboard}, we define the empirical checkerboard copula and introduce the nonparametric estimator of the lower tail copula based on this construction. All main results concerning the checkerboard approximation and tail copula processes are collected in section~\ref{sec:main}. Section~\ref{sec:des} develops the multiplier bootstrap procedures and establishes their asymptotic validity. Simulation studies illustrating the finite-sample performance of the proposed procedures are reported in section~\ref{sec:sim}. Our methodology based on empirical checkerboard copula and its bootstrap approximation is applied to two practical problems in section~\ref{sec:applications}. Proofs of all the requisite lemmas and main results are deferred to sections~\ref{sec:prooflemma} and~\ref{sec:ApA}, respectively.

\section{
Checkerboard Copula Construction and Tail Dependence}
\label{sec:checkerboard}
As mentioned earlier, the checkerboard copula provides a smooth, piecewise bilinear approximation of a copula by discretizing the unit square into a finite grid. It is particularly useful in settings where smoothness is required, for instance in empirical process theory and tail copula estimation.

\subsection{Population Checkerboard Copula}

Let $C$ be a copula on $[0,1]^2$.  
Fix an integer $m \ge 1$, called the checkerboard resolution or grid size. Now Lemma 2.3.5 of \citet{Nelsen2006} and definition 2.4  of \citet{GonzalezBarriosHoyosArguelles2020} will give us that, for every $(u,v)\in[0,1]^2$, there exist unique $i,j\in\{1,...,m\}$ such that $(u,v)\in I_{ij}:=\Big(\frac{i-1}{m},\frac{i}{m}\Big]\times \Big(\frac{j-1}{m},\frac{j}{m}\Big]$. When $i=1$ or $j=1$, we consider the intervals to be $[0,\frac1m]$. Now define,
\[
\mu(u):=\frac{u-\frac{(i-1)}{m}}{\frac{i}{m}-\frac{i-1}{m}}= m u - (i-1), \qquad
\mu(v) :=\frac{v-\frac{(j-1)}{m}}{\frac{j}{m}-\frac{j-1}{m}}= m v - (j-1).
\]
Thus the checkerboard copula associated with $C$ is defined as:
\[
C^{(m)}(u,v) := T_m(C)(u,v),
\]
where $T_m$ is the checkerboard approximation (piecewise bilinear interpolation) given by
\[
\begin{aligned}
T_m(C)(u,v)
&=
(1-\mu(u))(1-\mu(v))\, C\!\left(\tfrac{i-1}{m},\tfrac{j-1}{m}\right)
+ \mu(u)(1-\mu(v))\, C\!\left(\tfrac{i}{m},\tfrac{j-1}{m}\right) \\
&\quad
+ (1-\mu(u))\mu(v)\, C\!\left(\tfrac{i-1}{m},\tfrac{j}{m}\right)
+ \mu(u)\mu(v)\, C\!\left(\tfrac{i}{m},\tfrac{j}{m}\right),
\end{aligned}
\]
Therefore $T_m(C)$ is a valid copula (cf. Remark 2.3 of \citet{GonzalezBarriosHoyosArguelles2020}). The resulting function $C^{(m)}$ is continuous on $[0,1]^2$, bilinear on each rectangle $I_{ij}$, and is uniquely determined by the values of $C$ on the grid $\{0,1/m,\dots,1\}^2$.

\subsection{Empirical Checkerboard Copula}\label{sec:empcheckest}

Let $(U_1,V_1),\dots,(U_n,V_n)$ be pseudo-observations obtained from a bivariate sample with continuous margins, and let
\[
\hat C_n(u,v) := \frac{1}{n} \sum_{r=1}^n \mathbf{1}\{U_r \le u,\; V_r \le v\}
\]
denote the empirical copula. Then the empirical checkerboard copula is defined by 
\[
\hat C_n^{(m)} := T_m(\hat C_n),
\]
that is, the bilinear interpolation of the empirical copula on the $m \times m$ grid. The empirical checkerboard copula has been appeared several times in the literature; see for example  \citet{GonzalezBarriosHoyos2021}.
 Unlike $\hat C_n$, the checkerboard version $\hat C_n^{(m)}$ is continuous and avoids boundary discontinuities. Such smoothness is generally desirable in practical implementations. 

\subsection{Checkerboard-Based Tail Copula}

The lower tail copula associated with $C$ is defined as (cf.  \citet{Nelsen2006}):
\[
\Lambda_L(x,y)
=
\lim_{t \downarrow 0} \frac{C(tx,ty)}{t},
\qquad x,y \ge 0,
\]
whenever the limit exists everywhere on $\bar{\mathbb R}_{+}^2$. When $\Lambda_L$ exists, the lower tail dependence coefficient is defined by
\(
\lambda_L := \Lambda_L(1,1)\) (cf. \citet{Nelsen2006,schmidt2006}). 
Now following the definition of the estimator of $\Lambda_L$ based on classical empirical copula, we define the empirical checkerboard tail copula estimator of $\Lambda_L$ as:
\[
\widehat{\Lambda}_{L,n}^{(m)}(x,y)
:=
\frac{n}{k_n}
\hat C_n^{(m)}\!\left(\frac{k_n x}{n},\,\frac{k_n y}{n}\right),
\qquad x,y \ge 0,
\]
with some sequence $\{k_n\}_{n\geq 1}$. Subsequently, the checkerboard-based estimator of $\lambda_L$ is
\[
\hat\lambda_{L,n}^{(m)}
:=
\widehat{\Lambda}_{L,n}^{(m)}(1,1)
=
\frac{n}{k_n}
\hat C_n^{(m)}\!\left(\frac{k_n}{n},\,\frac{k_n}{n}\right).
\]
In the above definitions, $\frac{k_n}{n}$ denotes the proportion of samples used to construct the above lower tail estimator.



 An equivalent analogue of all the above definitions can be poised for upper tail copula. The upper tail copula associated with a copula $C$ is defined as:
\[
\Lambda_U(x,y)
=
\lim_{t \downarrow 0}
\frac{x t+y t-1 + C(1 - x t,\, 1 - y t)}{t},
\qquad x,y \ge 0,
\]
whenever the limit exists everywhere on $\bar{\mathbb R}_{+}^2$.  Analogous to lower tail counterpart, the empirical checkerboard upper tail copula estimator is defined by
\[
\widehat{\Lambda}_{U,n}^{(m)}(x,y)
:=
\frac{n}{k_n}
\Bigg[
\frac{k_n x}{n}
+ \frac{k_n y}{n}-1
+ \hat C_n^{(m)}\!\left(
1 - \frac{k_n x}{n},\,
1 - \frac{k_n y}{n}
\right)
\Bigg],
\qquad x,y \ge 0,
\]
and hence the checkerboard-based estimator of the upper tail dependence coefficient \(\lambda_U\) ($:= \Lambda_U(1,1)$) is given by
\[
\hat\lambda_{U,n}^{(m)}
:=
\widehat{\Lambda}_{U,n}^{(m)}(1,1)
=
\frac{n}{k_n}
\Bigg[
2\frac{k_n}{n}-1
+ \hat C_n^{(m)}\!\left(
1 - \frac{k_n}{n},\,
1 - \frac{k_n}{n}
\right)
\Bigg].
\]

\section{ Main Results}\label{sec:main}
In this section, we develop asymptotic theory for both the empirical checkerboard copula process and its tail counterpart. We divide this section into two subsections. We explore the consistency and weak convergence of the empirical checkerboard copula process in the first subsection. The second one extend the theory to the empirical checkerboard tail copula.
\subsection{Results on checkerboard copula process}
We begin this subsection by establishing the uniform consistency of the empirical
checkerboard copula estimator. This result provides the theoretical
foundation for the subsequent asymptotic analysis and justifies the use
of checkerboard smoothing as a valid approximation of the underlying
true copula.
\begin{theorem}[Consistency of Checkerboard estimator]\label{thm:consistency_checkerboard}
Let \(C\) be the true copula and  \(\widehat{C}_n^{(m)}\) be the
the empirical checkerboard copula estimator defined in section \ref{sec:empcheckest} where $m\equiv m_n\to\infty$ as $n\to\infty$.  Then,
\begin{equation}\label{eq:uniform_consistency}
\sup_{(u,v)\in[0,1]^2}
\bigl|
\widehat{C}_n^{(m)}(u,v) - C(u,v)
\bigr|
=
o(1),
\;\;\text{almost surely}\;\; \text{as } n\to\infty .
\end{equation}
\end{theorem}
If in addition to the assumptions of Theorem \ref{thm:consistency_checkerboard}, we also have $m/n^{1/2}\gg (\log \log n)^{-1/2}$, then `$o(1)$' in the RHS of the conclusion can be replaced by `$O\big[n^{-1/2}(\log\log n)^{1/2}\big]$'. In other words, under the additional condition on the checkerboard grid size the empirical checkerboard copula estimator can achieve the same convergence rate as the empirical copula estimator in almost sure sense. The proof of Theorem \ref{thm:consistency_checkerboard} is provided in section \ref{appAthm4.1}. The proof proceeds by decomposing the uniform error of the empirical
checkerboard copula into a stochastic and a deterministic component.
The stochastic term is controlled by expressing the checkerboard estimator
as the action of the corresponding linear interpolation operator \(T_m\) on the empirical
copula process. The operator \(T_m\) is shown to be norm--contractive on
\(\ell^\infty([0,1]^2)\), which allows to use the theory of the
empirical copula 
process (cf. \citet{janssen2012bernstein}). The deterministic bias term arises from the checkerboard approximation of
the underlying copula. This term is analyzed via Lipschitz property of the copula after fixing a particular grid. 
Combining the stochastic fluctuation and the deterministic approximation
error yield uniform almost sure consistency. 
Based on the uniform consistency of Theorem Theorem~\ref{thm:consistency_checkerboard}, we now explore the weak convergence of the empirical checkerboard copula process. We show that, under suitable smoothness
conditions, the checkerboard smoothing does not affect the first--order
asymptotic distribution of the empirical copula. In other words, the empirical checkerboard copula process
admits the same weak limit as the classical empirical copula process.

\begin{theorem}[Weak convergence of empirical checkerboard copula process.]\label{thm:weakconv_checkerboard} 
Let \(C\) be the true copula. If \(\widehat{C}_n^{(m)}\) is the empirical checkerboard copula estimator where $\sqrt{n}=o(m)$. Assume that
for each $j\in\{1,2\}$, the partial derivative
$\dot{C}_j$ of the true copula $C$ with respect to $j$th component exists and is continuous on
\[
\{(u_1,u_2)\in[0,1]^2:0<u_j<1\}.
\]
Then we have
\[
\sqrt{n}\bigl(\widehat{C}_n^{(m)} - C\bigr)
\;\rightsquigarrow\;
\mathbb{G}_C
\quad\text{in}\quad
\ell^\infty([0,1]^2),
\qquad \text{as}\;n\to\infty,
\]
where \(\mathbb{G}_C\) is the centered Gaussian process associated with the
empirical copula and admits the representation
\[
G_C(u_1,u_2)
=
\alpha_C(u_1,u_2)
-
\dot{C}_1(u_1,u_2)\,\alpha_C(u_1,1)
-
\dot{C}_2(u_1,u_2)\,\alpha_C(1,u_2),
\]
where $\alpha_C$ is a $C$-Brownian bridge, i.e., a centered Gaussian process with covariance
\[
\mathrm{cov}\big(\alpha_C(u_1,u_2), \alpha_C(u_1',u_2')\big)
=
C(u_1 \wedge u_1',\, u_2 \wedge u_2') - C(u_1,u_2)\,C(u_1',u_2').
\]
\end{theorem}
Note that if the grid size $m$ of the underlying checkerboard approximation is larger in order than $\sqrt{n}$, then the weak limit of the empirical checkerboard copula process matches with that of the empirical copula process. See for example Theorem 3 of \citet{fermanian2004weak}, Corollary 1 of \citet{Tsukahara2005} and Proposition 3.1 of \citet{Segers2012} for weak convergence of the empirical copula process. The smoothness condition on the true copula $C$, assumed in Theorem \ref{thm:weakconv_checkerboard}, is the same smoothness condition considered in \citet{Segers2012} for analyzing the empirical copula process. This condition is much weaker than the condition that the partial derivatives of the copula exist and are continuous on the entire unit square $[0,1]^2$, assumed for example in \citet{fermanian2004weak}, \citet{Tsukahara2005} and \citet{schmidt2006}. Such a relaxation is important since most of copula functions do not actually have continuous partial derivatives at $(0, 0)$ and $(1, 1)$. One can see Section 5 of \citet{Segers2012} for such examples. 

Similar to the proof of Theorem \ref{thm:consistency_checkerboard}, the primary step in establishing Theorem \ref{thm:weakconv_checkerboard} is to decompose the centered and scaled checkerboard copula estimator
into a stochastic fluctuation term and a deterministic bias term. 
The stochastic term is actually in terms of
the linear interpolation
operator \(T_m\) and the empirical copula process. Since \(T_m\) is a bounded linear operator on
\(\ell^\infty([0,1]^2)\), the extended continuous mapping theorem (cf. Lemma \ref{lem:5}) ensures that the weak
convergence of the empirical copula process is preserved under checkerboard smoothing. On the other hand, the deterministic bias term is asymptotically negligible when $\sqrt{n}=o(m)$. 
The weak
convergence of the empirical checkerboard copula then follows due to Slutsky’s
theorem. 

\subsection{Results on checkerboard tail copula process}\label{sec:tailcopula}
The asymptotic results of the empirical checkerboard copula process, established in the previous subsection ensure that the checkerboard smoothing
does not alter the first--order asymptotic behavior of the empirical copula
process under some growth rate of the grid size $m$. In this section, we extend these asymptotic properties to the checkerboard tail copula processes and the checkerboard tail dependence measures. First we study the strong consistency of the checkerboard tail copula processes.
\begin{theorem}[Strong consistency of the checkerboard tail copula]\label{thm:strongconstail} Under the conditions that $\frac{n}{\sqrt{k_n}}=o(m)$ and $(\log n)^2=o(k_n)$,
the empirical checkerboard tail copula satisfies
\[
\mathbf{P}\Big[\lim_{n\to\infty}d(\hat \Lambda_{L,n}^{(m)},\Lambda_L)=0\Big]=1  
\quad\text{and}\quad \mathbf{P}\Big[\lim_{n\to\infty}d(\hat \Lambda_{U,n}^{(m)},\Lambda_U)=0\Big]=1,
\]
where the metric $d(\cdot,\cdot)$ associated with the space $B_{\infty}(\bar{\mathbb{R}}_{+}^2)$ is defined in section \ref{sec:intro}.
\end{theorem}
The proof of Theorem \ref{thm:strongconstail} is given in section \ref{appAthm4.5}. The proof relies on the application of the strong consistency of the empirical tail copula process established as Theorem~6 in \citet{schmidt2006} and to utilize Theorem \ref{eq:uniform_consistency}. The application of Theorem~6 of \citet{schmidt2006} additionally requires $(\log n)^2 = o(k_n)$. The condition $\frac{n}{\sqrt{k_n}} = o(m)$ is essential in order to effectively apply Theorem \ref{eq:uniform_consistency}. The next result is on the weak convergence of the checkerboard tail copula processes for which we require following two regularity conditions:
\begin{enumerate}
\item[(C.1.L)] The first-order partial derivatives of $\Lambda_L(x,y)$ satisfy the condition:
\[
\partial_x \Lambda_L,\;\partial_y \Lambda_L \text{ exist and are continuous on }
\left\{ (x,y) \in \overline{\mathbb{R}}_+^2 \,\middle|\, 0 < x,y < \infty \right\}.
\]
\item[(C.1.U)] Moreover, we define $\partial_x \Lambda_L=0=\partial_y \Lambda_L$ on the set
\(\left\{ (x,y) \in \overline{\mathbb{R}}_+^2 \,\middle|\, x,y \in \{0,\infty\} \right\}.\) 

The first-order partial derivatives of $\Lambda_L(x,y)$ satisfy the condition:
\[
\partial_x \Lambda_L,\;\partial_y \Lambda_L \text{ exist and are continuous on }
\left\{ (x,y) \in \overline{\mathbb{R}}_+^2 \,\middle|\, 0 < x,y < \infty \right\}.
\]
Moreover, we define $\partial_x \Lambda_L=0=\partial_y \Lambda_L$ on the set
\(\left\{ (x,y) \in \overline{\mathbb{R}}_+^2 \,\middle|\, x,y \in \{0,\infty\} \right\}.\)

\item[(C.2.L)] There exists a function $A:\mathbb{R}_+ \to \mathbb{R}_+$ such that $A(t)\to 0$ as $t\to\infty$ and the lower tail copula $\Lambda_L(x,y)$ satisfies: 
\[
\big|\Lambda_L(x,y) - t\,C(x/t, y/t)\big| = O\big(A(t)\big),
\qquad t \to \infty,
\]
locally uniformly for $(x,y) \in \overline{\mathbb{R}}_+^2$.
\item[(C.2.U)] A similar version for upper tail can be provided. There exists a function $A_U:\mathbb{R}_+\to\mathbb{R}_+$ such that
$A_U(t)\to 0$ as $t\to\infty$ and
\[
\Bigg|\Lambda_U(x,y)
-
t\Big[
\frac{x}{t}+\frac{y}{t}-1
+
C\!\Big(1-\frac{x}{t},\,1-\frac{y}{t}\Big)
\Big]\Bigg|=O(
A_U(t)),
\]
locally uniformly for $(x,y)\in\overline{\mathbb{R}}_+^2$.
\end{enumerate}
We denote the regularity conditions by (C.1) and (C.2) only. For lower (or upper) tail these conditions simply mean (C.1.L) and (C.2.L) (or (C.1.U) and (C.2.U)). The condition (C.1) 
is necessary for a tail functional to admit a first-order expansion 
and then to apply the functional delta method to obtain the limit process. This regularity condition is generally indispensable in analyzing the tail copula process since without it one can not have hadamard directional derivative of the underlying true copula $C$ with respect to some suitable set, required for the application of functional Delta method; see for example \citet{BucherDette2013}, \citet{Kiriliouk2018,Bormann2020}. The continuous partial derivative of $C$ in the interior of $[0, 1]^2$, assumed in Theorem \ref{thm:weakconv_checkerboard},  does not generally imply the condition (C.1). Continuity of the partial derivatives of the true copula $C$ at $(0, 0)$ and the uniform convergence of those partial derivatives are also required. However, the continuity of the partial derivative of $C$ at $(0, 0)$ does not generally hold as mentioned in the previous section, although the condition (C.1) is quite universal. See Remark \ref{rem:examples} for two examples where partial derivative of $C$ is not continuous at $(0, 0)$ but condition (C.1) holds.  The condition (C.2) 
provides a control on the bias in the tail approximation via a second-order function $A(t)$, which vanishes as $t\to\infty$. This is required for obtaining stochastic expansions of $\big(\widehat{\Lambda}_{L,n}^{(m)} - \Lambda_L\big)$ and to control the bias term present in the expansion relative to the stochastic fluctuation of the leading term. The condition (C.2) is a relaxation of the second-order condition usually assumed in extreme value theory (cf. \citet{Hall1982, deHaanFerreira2006}) and is routinely used in the literature on copula (see \citet{BucherDette2013}, \citet{Bormann2020}). Now we are ready to state the weak convergence of the checkerboard tail copula.
\begin{theorem}[Weak convergence of checkerboard-based tail copula process]\label{thm:weakconv_tail_checkerboard}
Suppose that conditions (C.1) and (C.2) are true. Then provided, $k_n=o(n)$ with $\sqrt{k_n}A(n/k_n)\to 0$ and $\frac{n}{\sqrt{k_n}}=o(m)$, we have; 
\begin{align*}
&\sqrt{k_n}
\Big(
\widehat{\Lambda}_{L,n}^{(m)} - \Lambda_L
\Big)
\;\rightsquigarrow\;
\mathbb{G}_{\hat{\Lambda}_L},\quad\text{in}\; B_\infty(\overline{\mathbb{R}}_+^2),
\qquad n\to\infty,\\
& \sqrt{k_n}
\Big(
\widehat{\Lambda}_{U,n}^{(m)} - \Lambda_U
\Big)
\;\rightsquigarrow\;
\mathbb{G}_{\hat{\Lambda}_U},
\quad\text{in } \;B_\infty(\overline{\mathbb{R}}_+^2),
\qquad n\to\infty,
\end{align*}
where \(\mathbb{G}_{\hat{\Lambda}_L}\) and \(\mathbb{G}_{\hat{\Lambda}_U}\) can be expressed as
\begin{align*}
&\mathbb{G}_{\hat{\Lambda}_L}(x,y) = \mathbb{G}_{\hat{\Lambda}_L^*}(x,y)
- \frac{\partial}{\partial x} \Lambda_L(x,y)\, \mathbb{G}_{\hat{\Lambda}_L^*}(x,\infty) 
- \frac{\partial}{\partial y} \Lambda_L(x,y)\, \mathbb{G}_{\hat{\Lambda}_L^*}(\infty,y),\\
&\mathbb{G}_{\hat{\Lambda}_U}(x,y)
=
\mathbb{G}_{\hat{\Lambda}_U^*}(x,y)
-
\frac{\partial}{\partial x}\Lambda_U(x,y)\,
\mathbb{G}_{\hat{\Lambda}_U^*}(x,\infty)
-
\frac{\partial}{\partial y}\Lambda_U(x,y)\,
\mathbb{G}_{\hat{\Lambda}_U^*}(\infty,y).
\end{align*}
Here $\mathbb{G}_{\hat{\Lambda}_L^*}(x,y)$ is a  centered tight Gaussian process with the covariance structure given by
\begin{align*}
&\mathbb{E}\Big[ \mathbb{G}_{\hat{\Lambda}_L^*}(x,y)\, \mathbb{G}_{\hat{\Lambda}_L^*}(\bar x, \bar y) \Big] 
= \Lambda_L\big( \min\{x, \bar x\}, \min\{y, \bar y\} \big),
\quad (x,y), (\bar x, \bar y) \in \overline{\mathbb{R}}_+^2.
\end{align*}
The same holds for $\mathbb{G}_{\hat{\Lambda}_U^*}(x,y)$ replacing $\Lambda_L$ by $\Lambda_U$.
\end{theorem}
Note that the limiting processes are same as that in case of empirical tail copula process. First we show that the deviation of the checkerboard tail copula process from the empirical tail copula process is of order $o(k_n^{-1/2})$, uniformly over $[0, 1]^2$. Then we utilize the weak convergence of the empirical tail copula process (cf. Theorem~2.2 of \citet{BucherDette2013}) in order to establish Theorem \ref{thm:weakconv_tail_checkerboard}. The proof is presented in Section \ref{appAthm4.3}.
The Gaussian limit of the tail
dependence coefficients $\hat \lambda_{L,n}^{(m)}$ and $\hat \lambda_{U,n}^{(m)}$ follow directly from the weak convergence of the empirical checkerboard tail copula process established in
Theorem~\ref{thm:weakconv_tail_checkerboard}. We state the result for the lower tail dependence coefficient as the following corollary. The result for the upper tail coefficient is analogous.
\begin{corollary}[Weak convergence of checkerboard-based tail coefficient.]\label{cor:asymp_normal_tailcoeff}
Consider the same set-up of Theorem \ref{thm:weakconv_tail_checkerboard}. Then,
\[
\sqrt{k_n}\,
\bigl(\hat \lambda_{L,n}^{(m)} - \lambda_L\bigr)
\;\overset{d}{\longrightarrow}\;
N\bigl(0, \sigma_L^2\bigr),
\qquad n\to\infty,
\]
where the asymptotic variance \(\sigma_L^2\) is given by
\[
\sigma_L^2
=
\lambda_L
+
\Bigg(\frac{\partial}{\partial x} \Lambda_L(1,1)\Bigg)^2
+
\Bigg(\frac{\partial}{\partial y} \Lambda_L(1,1)\Bigg)^2
+
2\,\lambda_L
\Bigg[
\Bigg(\frac{\partial}{\partial x} \Lambda_L(1,1)-1\Bigg)
\Bigg(\frac{\partial}{\partial y} \Lambda_L(1,1)-1\Bigg)
-1
\Bigg].
\]
The above convergence holds also for $\hat \lambda_{U, n}^{(m)}$ if we replace $\Lambda_L$ by $\Lambda_U$.
\end{corollary}
Under the assumed regularity conditions, the corollary follows from  Theorem \ref{thm:weakconv_tail_checkerboard} by plugging in $(x,y)=(1,1)$. This result justifies the use of checkerboard smoothing in finite samples
while retaining valid asymptotic inference, and it enables the construction
of asymptotically valid confidence intervals and hypothesis tests for
\(\lambda_L\) based on the normal approximations.
\begin{remark}\label{rem:examples}
The following two examples are among many forms of copulas where the partial derivative of $C$ is not continuous at $(0, 0)$ but the condition (C.1) is true.
\begin{enumerate}
\item[(i)] \textbf{Gaussian copula.} 
Example~5.1 of \citet{Segers2012} argued that Gaussian copula does not have continuous partial derivative at $(0, 0)$. However, condition (C.1) is true. Since it is asymptotically independent in the lower tail (see, e.g., \citet{Joe2014,Nelsen2006}) for $|\rho|<1$, its lower tail copula satisfies
\[
\Lambda_L(x,y)\equiv 0,\qquad (x,y)\in\overline{\mathbb{R}}_+^2.
\]
Hence
\[
\partial_x\Lambda_L=\partial_y\Lambda_L\equiv 0,
\]
which are continuous on $(0,\infty)^2$. Therefore, (C.1.L) holds trivially.

\item[(ii)] \textbf{Clayton copula.}
Example~5.2 of \citet{Segers2012} showed that some choices of copula in the Archimedean family may not have partial derivative at $(0, 0)$ or at $(1, 1)$. However, the regularity condition (C.1) may still hold. Let us consider the Clayton copula in that regard. 
Note that the Clayton copula is given by
\[C_\theta(u,v)
=
\left(
u^{-\theta}
+
v^{-\theta}
-
1
\right)^{-1/\theta},
\quad
\theta\in(0,\infty),
\quad
(u,v)\in[0,1]^2.\]. Its lower tail copula is
\[
\Lambda_L(x,y)
=
\left(x^{-\theta}+y^{-\theta}\right)^{-1/\theta},
\qquad \theta>0.
\]
Consequently,
\[
\partial_x\Lambda_L(x,y)
=
x^{-\theta-1}
\left(x^{-\theta}+y^{-\theta}\right)^{-1/\theta-1},
\]
and similarly for $\partial_y\Lambda_L$. Both derivatives are continuous on
$(0,\infty)^2$, and with the convention
\[
\partial_x\Lambda_L=\partial_y\Lambda_L=0
\quad\text{on}\quad
\{(x,y):x,y\in\{0,\infty\}\},
\]
Thus regularity condition (C.1.L) holds. Similarly, one can show that the condition (C.1.U) also holds since $\Lambda_U(x,y)\equiv 0,\; (x,y)\in\overline{\mathbb{R}}_+^2$.
\end{enumerate}
\end{remark}

\section{Bootstrap approximation}\label{sec:des}
In the previous section, we establish that the empirical checkerboard copula process and its tail counterpart both converge to centered tight Gaussian processes.  
However, the covariance structure in either case depends on the unknown copula or tail copula and its partial derivatives. Hence the limiting processes are analytically intractable and may not in general be used for drawing inference in practice. This motivates the construction of a bootstrap procedure
that consistently mimics the distribution of the underlying empirical checkerboard copula and tail copula processes without requiring to estimate the unknowns present in the limit Gaussian processes. 
We adapt the idea of multiplier bootstrap of \citet{Kosorok2008} and  \citet{BucherDette2013}, 
developed in case of the empirical copula process. Although the empirical bootstrap of \citet{Efron1979} works for approximating the distribution of the empirical copula process, it is ill-suited for the empirical tail copula process. This is because the effective sample size in the tail region is of order $k_n \ll n$, and resampling from the full empirical distribution does not preserve the extreme-value structure required for valid tail estimation; see for example \citet{BucherDette2010}. Multiplier bootstrap methods, in contrast, operate by reweighting the original observations and
are known to be particularly effective for empirical process–type limits (cf. \citet{Kosorok2008}). It allows one to replicate the asymptotic fluctuations both in the middle as well as in the tails of the underlying distribution of interest. We divide this section in two parts. In the first subsection, we develop and justify the multiplier bootstrap approximation of the distribution of the empirical checkerboard copula process. In the second one, we extend the notion of multiplier bootstrap to the checkerboard tail copula.
\subsection{Bootstrap for empirical checkerboard copula}
Let $\xi_1,\dots,\xi_n$ be i.i.d.\ positive sub-exponential random variables, independent of the data, such that $\mathbf E(\xi_1)=\mu\in(0,\infty),\;$ and $\mathrm{Var}(\xi_1)=\tau^2\in(0,\infty).$ 
When $\mu =\tau$, then the resulting multiplier bootstrap is sometime termed as the perturbation bootstrap (cf. \citet{Jin2001,Choudhury2026}). Define the weighted empirical joint distribution function by
\[
H_n^\xi(x,y)
=
\frac{1}{n}\sum_{i=1}^n
\frac{\xi_i}{\bar\xi_n}
\mathbf 1\{X_{i}\le x,\; Y_{i}\le y\},
\]
and the corresponding weighted marginal distributions by
\[
F_{n}^\xi(x)
=
\frac{1}{n}\sum_{i=1}^n
\frac{\xi_i}{\bar\xi_n}
\mathbf 1\{X_{i}\le x\}\quad\text{and}
\quad G_{n}^\xi(y)
=
\frac{1}{n}\sum_{i=1}^n
\frac{\xi_i}{\bar\xi_n}
\mathbf 1\{Y_{i}\le y\},
\]
where $\bar{\xi}_n$ is the sample mean of $\xi_1, \dots, \xi_n$. 
In the above, the normalization by $\bar\xi_n$ is essential 
to ensure that $H_n^\xi$, $F_n^\xi$ and $G_n^\xi$ are all well defined distribution functions. 
Based on the aforementioned weighted empirical distributions, define the bootstrapped empirical copula as
\[
\hat{C}_n^{\xi,\xi}(u,v)
=
H_n^\xi\!\left(
(F_{n}^\xi)^{-}(u),
(G_{n}^\xi)^{-}(v)
\right),
\]
where for any real distribution function $K$, we denote its left continuous generalized inverse as,
\[
K^{-}(p) := 
\begin{cases}
\inf \{x \in \mathbb R : K(x) \ge p\}, & 0 < p \le 1, \\
\sup \{x \in \mathbb R : K(x) = 0\}, & p = 0.
\end{cases}
\]
 Subsequently define the bootstrapped empirical checkerboard empirical copula as
\begin{align}\label{weqn:mbec}
\widehat C_n^{(m),\xi,\xi}(u, v)
=
T_m\bigl(\hat C_n^{\xi,\xi}(u, v)\bigr),
\end{align}
and the corresponding centered version as
\(\sqrt{n}\big(\hat C_n^{(m),\xi,\xi}-\hat{C}_n^{(m)}\big)\).
Here and always $T_m$ denotes the empirical checkerboard operator as defined in
Section~\ref{sec:checkerboard} and $\hat{C}_n^{(m)}$ is the empirical checkerboard copula. 
We explore the validity of the multiplier bootstrap approximation of the distribution of the actual empirical checkerboard copula  in the following theorem which is the bootstrap analogue of Theorem \ref{thm:weakconv_checkerboard}.
\begin{theorem}[Bootstrap consistency of the checkerboard-based copula]\label{thm:bootcopula} Under the set-up of Theorem \ref{thm:weakconv_checkerboard} we have
\[
\frac{\mu}{\tau}\sqrt{n}\Big(\hat C_n^{(m),\xi,\xi}-\hat{C}_n^{(m)}\Big)\overset{\mathbf P}{\underset{\xi}{\rightsquigarrow}}\;\ 
\mathbb G_{C}
\quad
\text{in } \ell^\infty([0,1]^2),
\]
where $\mathbb{G}_C$ is the centered tight Gaussian process as defined in Theorem \ref{thm:weakconv_checkerboard}.
\end{theorem}
The above theorem essentially tells us that one can approximate the distribution of the centered and scaled empirical checkerboard copula process  $\sqrt{n}\big(\hat{C}_n^{(m)} - C\big)$ by the conditional distribution (given the observed samples) of the bootstrap counterpart $\frac{\mu}{\tau}\sqrt{n}\big(\hat C_n^{(m),\xi,\xi}-\hat{C}_n^{(m)}\big)$. Hence inference can be drawn about the true copula $C$ based on the empirical checkerboard copula $\hat{C}_n^{(m)}$ without requiring one to estimate the unknowns present in the weak limit $\mathbb{G}_C$ of  $\sqrt{n}\big(\hat{C}_n^{(m)} - C\big)$. The factor $\mu/\tau$ present in Theorem \ref{thm:bootcopula} is required to standardize the multipliers so that the conditional covariance structure matches with the original covariance structure asymptotically. Thus one can simply consider $\mu = \tau$ and get rid of this standardization term.
The proof of Theorem \ref{thm:bootcopula} is provided in section \ref{sec:klj}. The primary step is to write
\[
\frac{\mu}{\tau}\sqrt{n}\big(\hat C_n^{(m),\xi,\xi}-\hat{C}_n^{(m)}\big) = \frac{\mu}{\tau}\sqrt{n}\Big(\hat C_n^{\xi,\xi}-\hat{C}_n\Big) + R_n,
\]
where $R_n$ is an error term. We utilize Theorem 2.4 of \citet{Bucher2011} to obtain the weak convergence of the first term. 
Then we show that the error term $R_n$ is of order $O_p(m^{-1} + (\log n)/n)$, uniformly on any compact set, by exploiting the properties of the bilinear checkerboard operator $T_m$, properties of the left continuous generalized inverse of a distribution function and the sub-exponential bound  of the bootstrap multipliers. 
Theorem \ref{thm:bootcopula} then follows by Slutsky’s theorem. The handling of the error term $R_n$ tells us that one can also center $\hat C_n^{(m),\xi,\xi}$ by the empirical copula $\hat C_n$ instead of the empirical checkerboard copula $\hat{C}_n^{(m)}$.

\subsection{Bootstrap for empirical checkerboard tail copula}\label{sec:des1}
In this section, we explore whether multiplier bootstrap approximation of the empirical checkerboard copula can be extended to the corresponding tail copula. Recall the definition of the bootstrapped empirical checkerboard copula as in (\ref{weqn:mbec}) and define 
the corresponding bootstrap estimator of the checkerboard based lower tail copula
\[
\widehat{\Lambda}^{(m),\xi,\xi}_{L,n}(x,y)
=
\frac{n}{k_n}\,
\widehat C_n^{(m),\xi,\xi}
\!\left(
\frac{k_n x}{n},
\frac{k_n y}{n}
\right),
\qquad (x,y)\in\overline{\mathbb R}_+^2,
\]
and the upper tail copula
\[
\widehat{\Lambda}_{U,n}^{(m),\xi,\xi}(x,y)
:=
\frac{n}{k_n}
\Bigg[
\frac{k_n x}{n}
+ \frac{k_n y}{n}-1
+ \hat C_n^{(m),\xi,\xi}\!\left(
1 - \frac{k_n x}{n},\,
1 - \frac{k_n y}{n}
\right)
\Bigg],
\quad (x,y)\in\overline{\mathbb R}_+^2.
\]
Subsequently, the bootstrapped checkerboard tail copula processes after proper centering and scaling are given by
\[
\alpha_{L, n}^{(m)}(x,y)
=
\frac{\mu}{\tau}\sqrt{k_n}\,
\Bigl(
\widehat{\Lambda}^{(m),\xi,\xi}_{L,n}(x,y)
-
\widehat{\Lambda}^{(m)}_{L,n}(x,y)
\Bigr),
\]
\[
\alpha_{U, n}^{(m)}(x,y)
=
\frac{\mu}{\tau}\sqrt{k_n}\,
\Bigl(
\widehat{\Lambda}^{(m),\xi,\xi}_{U,n}(x,y)
-
\widehat{\Lambda}^{(m)}_{U,n}(x,y)
\Bigr).
\]
We now show that the conditional distribution of the bootstrap processes $\alpha_{L, n}^{(m)}$ and $\alpha_{U, n}^{(m)}$ can effectively approximate the distribution of the centered and scaled lower and upper tail empirical checkerboard copula processes respectively. We state the result as the next theorem.
\begin{theorem}[Bootstrap consistency of the checkerboard-based tail copula]
\label{thm:cb-dm-bootstrap}
Consider the setup of Theorem \ref{thm:weakconv_tail_checkerboard}. Additionally assume that $k_n >> (\log n)^2$.
Then, conditionally on the data, we have
\[
\alpha_{L, n}^{(m)}
\ \overset{\mathbf P}{\underset{\xi}{\rightsquigarrow}}\;\ 
\mathbb G_{\hat\Lambda_L}\;\;\text{and}\;\;\alpha_{U, n}^{(m)}
\ \overset{\mathbf P}{\underset{\xi}{\rightsquigarrow}}\;\ 
\mathbb G_{\hat\Lambda_u}
\quad
\text{in } B_\infty(\overline{\mathbb R}_+^2),
\]
where
$\mathbb G_{\hat \Lambda_L}$ and $\mathbb G_{\hat\Lambda_u}$ are the centered tight Gaussian process 
\end{theorem}
The proof of Theorem \ref{thm:cb-dm-bootstrap} is presented in section \ref{appAthm4.6}. The proof is similar to that of Theorem \ref{thm:bootcopula}. 
Theorem \ref{thm:cb-dm-bootstrap} indicates that the multiplier bootstrap can be used to  construct confidence regions and hypothesis
tests for some feature of the underlying tail copulas $\Lambda_L$ and $\Lambda_U$. One does not require to explicit estimation of the unknowns present in the weak limits $\mathbb{G}_{\hat{\Lambda}_L}$ and $\mathbb{G}_{\hat{\Lambda}_U}$, present in Theorem \ref{thm:weakconv_tail_checkerboard}. Important features of the tail copulas are the tail dependent coefficients $\lambda_L$ and $\lambda_U$. We have the following corollary on the bootstrap approximation of the distribution of the tail dependence coefficients as an immediate consequence of Theorem \ref{thm:cb-dm-bootstrap}. 
\begin{corollary}[Bootstrap consistency of tail copula coefficients]\label{cor:bootstrap_lambdaL}
Consider the same set-up of Theorem \ref{thm:cb-dm-bootstrap}. If for $j= L, U$,
\[
\lambda_j = \Lambda_j(1,1), 
\qquad
\hat{\lambda}_{j,n}^{(m)} = \hat{\Lambda}_{j,n}^{(m)}(1,1),
\qquad
\hat{\lambda}_{j,n}^{(m),\xi,\xi} = \hat{\Lambda}_{j,n}^{(m),\xi,\xi}(1,1),
\]
then we have
\[\sup_{x\in \mathbb{R}}\Big|\mathbf{P}_{\xi}\Big(\frac{\mu}{\tau}\sqrt{k_n}\Big(
\hat{\lambda}_{j,n}^{(m),\xi,\xi}
-
\hat{\lambda}_{j,n}^{(m)}
\Big) \leq x\Big) - \mathbf{P}\Big(\sqrt{k_n}\Big(
\hat{\lambda}_{j,n}^{(m)}
-
\hat{\lambda}_{j}
\Big) \leq x\Big)\Big|
\xrightarrow{\mathbf{P}} 0, \; \text{as}\; n \rightarrow \infty.
\] 
Here $\mathbf{P}_{\xi}$ denotes the conditional probability of the bootstrap multipliers given the data.
\end{corollary}

\section{Simulation study}\label{sec:sim}
In this section, we only investigate the finite--sample performance of the proposed
checkerboard tail copula estimator and the validity of the associated multiplier bootstrap (cf. Theorem \ref{thm:cb-dm-bootstrap}). The simulations are designed to assess (i) the effect of
checkerboard smoothing on estimation accuracy and (ii) the reliability of
bootstrap inference for the lower (and/or upper) tail dependence coefficient in the
unknown--marginal setting, where asymptotic inference is infeasible.

\subsection{Data generating models}

We consider bivariate samples $(X_i,Y_i)_{i=1}^n$ generated from copula models
with known lower tail dependence structure. In particular, we focus on the
Clayton copula with parameter $\theta>0$, for which the lower tail dependence
coefficient is given by $\lambda_L = 2^{-1/\theta}$, and on the Gumbel copula with parameter $\theta>0$, for which the upper tail dependence
coefficient is given by $\lambda_U = 2-2^{-1/\theta}$. In all scenarios, the marginal distributions are assumed to be unknown. This setup reflects the conditions under which the theoretical
results of Sections~\ref{sec:main}--\ref{sec:des} are derived and
ensures that the additional stochastic variability induced by marginal
estimation is present.

\subsection{Estimators and tuning parameters}

We compare the classical empirical tail copula estimator
$\hat{\Lambda}_{L,n}$ with the proposed checkerboard--smoothed estimator
$\hat{\Lambda}_{L,n}^{(m)}$. The lower tail dependence coefficient is estimated
by point evaluation at $(1,1)$. Sample sizes $n\in\{500,1000,2000\}$ are considered. The tail fraction is chosen
as $k_n=\lfloor n^{\alpha}\rfloor$  while the
checkerboard resolution is set to $m=\lfloor n^{\beta}\rfloor$ with  sufficiently varying choices of $(\alpha,\beta)$ as given by, $(\alpha,\beta)\in\{(0.75,0.75),\;(0.8,0.85),\;(0.9,0.95)\}$ for lower tail dependence and $(\alpha,\beta)\in\{(0.6,0.95),\;(0.7,0.90),\;(0.8,0.85)\}$ for upper tail dependence.

\subsection{Bootstrap implementation}

Inference is conducted using the proposed checkerboard--based multiplier
bootstrap. The multiplier variables are generated independently from a standard
exponential distribution and normalized by their sample mean. The finite sample results are based on $1000$ Monte Carlo replications. For each Monte Carlo replication, $B=500$ bootstrap samples are generated. 

\subsection{Performance measures}

Estimation accuracy is evaluated in terms of empirical bias and mean
squared error of the checkerboard estimator and classical empirical estimator of the lower and upper tail dependence coefficient.
Inference performance is assessed by empirical coverage probabilities and
average lengths of nominal $90\%$ confidence intervals constructed using the proposed bootstrap method. Let us denote $P_n:=\frac{\mu}{\tau}\sqrt{k_n}\Big(
\hat{\lambda}_{L,n}^{(m),\xi,\xi}
-
\hat{\lambda}_{L,n}^{(m)}
\Big)$ and $(P_n)_{\gamma}$ be the $\gamma$-th quantile of the bootstrap distribution of $P_n$. Then $100(1-\gamma)\%$ both-sided Bootstrap percentile interval for $\lambda_L$ is given by: $\Big[\hat{\lambda}_{L,n}^{(m)}-\frac{\tau}{\mu}\frac{\big(P_n\big)_{1-\frac{\gamma}{2}}}{k_n^{1/2}},\;\hat{\lambda}_{L,n}^{(m)}-\frac{\tau}{\mu}\frac{\big(P_n\big)_{\frac{\gamma}{2}}}{k_n^{1/2}}\Big].$ Similarly we can define this for upper tail coefficient $\lambda_U$.
\begin{table}[!ht]
\centering
\caption{Empirical bias and mean squared error (MSE) of the lower tail dependence
coefficient estimator for the Clayton copula with $\theta=2$. Results are based on 1000 Monte
Carlo replications. True $\lambda_L=0.707$}
\label{tab:clayton_bias_mse}
\begin{tabular}{cccccc}
\hline
$n$ & $(\alpha,\beta)$  &
$\text{CB}\big(\hat{\lambda}_{L,n}^{(m)}\big)$ &$\text{EB}\big(\hat{\lambda}_{L,n}\big)$ &
$\text{CMSE}\big(\hat{\lambda}_{L,n}^{(m)}\big)$ & $\text{EMSE}\big(\hat{\lambda}_{L,n}\big)$ \\
\hline
500  & (0.75,0.75) & 0.0241 & 0.0967 &0.0021 & 0.0345  \\[2pt]
     & (0.80,0.85) & 0.0180 & 0.1036 & 0.0017 & 0.0461  \\[2pt]
     & (0.90,0.95) & 0.0152 & 0.0877 & 0.0011 & 0.0185  \\
\hline
1000 & (0.75,0.75) & 0.0102 & 0.0776 & 0.00086 & 0.0094  \\[2pt]
     & (0.80,0.85) & 0.0092& 0.0744  & 0.00061 & 0.0083\\[2pt]
     & (0.90,0.95) & 0.0096& 0.0532 & 0.00058 & 0.0066  \\
\hline
2000 & (0.75,0.75) & 0.0052 &0.0392 &  0.00018 & 0.0045 \\[2pt]
     & (0.80,0.85) & 0.0032 & 0.0221 & 0.00011 & 0.0022  \\[2pt]
     & (0.90,0.95) & 0.0024 & 0.0108 & 0.00009 & 0.0014 \\
\hline
\end{tabular}
\\[0.5ex]
\parbox{0.95\linewidth}{\footnotesize
CB: Bias of Checkerboard Estimator $\hat{\lambda}_{L,n}^{(m)}$;
EB: Bias of Classical Empirical Estimator $\hat{\lambda}_{L,n}$;
CMSE: Mean Squared Error of Checkerboard Estimator $\hat{\lambda}_{L,n}^{(m)}$;
EMSE: Mean Squared Error of Classical Empirical Estimator $\hat{\lambda}_{L,n}$.
}
\end{table}
Table~\ref{tab:clayton_bias_mse} reports the empirical bias and MSE of the lower tail dependence coefficient estimators for the Clayton copula with $\theta = 2$ $(\lambda_L \approx 0.707)$. Across all sample sizes and tuning parameter choices, the checkerboard estimator $\hat{\lambda}{L,n}^{(m)}$ consistently outperforms the classical empirical estimator $\hat{\lambda}_{L,n}$ in terms of both bias and MSE. The improvement is particularly pronounced for smaller samples. For example, when $n=500$, the checkerboard estimator substantially reduces the upward bias and achieves MSE values that are more than an order of magnitude smaller than those of the classical estimator. As $n$ increases, both estimators improve, but the checkerboard estimator remains uniformly superior, exhibiting faster decay in both bias and MSE. These findings demonstrate the effectiveness of checkerboard smoothing for stable and efficient lower tail dependence estimation.

\begin{table}[!ht]
\centering
\caption{Empirical bias and mean squared error (MSE) of the upper tail dependence
coefficient estimator for the Gumbel copula with $\theta=2$. Results are based on 1000 Monte
Carlo replications. True $\lambda_U=0.586$}
\label{tab:gumbel_bias_mse}
\begin{tabular}{cccccc}
\hline
$n$ & $(\alpha,\beta)$ &
$\text{CB}\big(\hat{\lambda}_{U,n}^{(m)}\big)$ & $\text{EB}\big(\hat{\lambda}_{U,n}\big)$ &
$\text{CMSE}\big(\hat{\lambda}_{U,n}^{(m)}\big)$ & $\text{EMSE}\big(\hat{\lambda}_{U,n}\big)$ \\
\hline
500  & (0.60,0.95) & 0.0162 & 0.0884 & 0.0034 &  0.0721  \\[2pt]
     & (0.70,0.90) & 0.0115 & 0.0901 & 0.0021 & 0.0697  \\[2pt]
     & (0.80,0.85) & 0.0206 & 0.1041 & 0.0037 & 0.0843    \\
\hline
1000 & (0.60,0.95) & 0.0076 & 0.0541 & 0.00092 & 0.0342 \\[2pt]
     & (0.70,0.90) & 0.0062 & 0.0210 & 0.00066  & 0.0120 \\[2pt]
     & (0.80,0.85) & 0.0093 & 0.0268 & 0.00102 & 0.0226 \\
\hline
2000 & (0.60,0.95) & 0.0042 & 0.0163 & 0.00043 & 0.0092 \\[2pt]
     & (0.70,0.90) & 0.0016 & 0.0098  & 0.00021 & 0.0082 \\[2pt]
     & (0.80,0.85) & 0.0057 & 0.0105  & 0.00089 & 0.0095 \\
\hline
\end{tabular}
\\[0.5ex]
\parbox{0.95\linewidth}{\footnotesize
CB: Bias of Checkerboard Estimator $\hat{\lambda}_{U,n}^{(m)}$;
EB: Bias of Classical Empirical Estimator $\hat{\lambda}_{U,n}$;
CMSE: Mean Squared Error of Checkerboard Estimator $\hat{\lambda}_{U,n}^{(m)}$;
EMSE: Mean Squared Error of Classical Empirical Estimator $\hat{\lambda}_{U,n}$.
}
\end{table}

Table~\ref{tab:gumbel_bias_mse} shows that the checkerboard estimator $\hat{\lambda}_{U,n}^{(m)}$ consistently outperforms the classical empirical estimator $\hat{\lambda}_{U,n}$ for estimating the upper tail dependence coefficient of the Gumbel copula $(\lambda_U \approx 0.586)$. Across all sample sizes and tuning parameter choices, the checkerboard estimator exhibits smaller bias and substantially lower MSE, with the gains being most pronounced for smaller samples. Although both estimators improve as $n$ increases, the checkerboard estimator maintains a clear advantage, achieving faster reduction in both bias and MSE. Intermediate tuning choices such as $(\alpha,\beta)=(0.70,0.90)$ generally provide the best balance between bias and variance. Overall, the results demonstrate that checkerboard smoothing substantially improves the finite-sample accuracy and stability of upper tail dependence estimation.
\begin{table}[!ht]
\centering
\caption{Empirical coverage probabilities and average width of nominal
90\% bootstrap confidence intervals for the lower tail dependence coefficient
(Clayton copula). True $\lambda_L=0.707$}
\label{tab:clayton_bootstrap}
\begin{tabular}{cccccc}
\hline
$n$ & $(\alpha,\beta)$ &
Coverage & Coverage &
Average width & Average width \\
 &  & 
(Checkerboard) & (Classical) & (Checkerboard) & (Classical) \\
\hline
500  & (0.75,0.75) & 0.864 & 0.804 &  0.197 & 0.343 \\[2pt]
     & (0.80,0.85) & 0.874 & 0.818 & 0.192 & 0.402 \\[2pt]
     & (0.90,0.95) & 0.872 & 0.822 & 0.186 & 0.304 \\
\hline
1000 & (0.75,0.75) & 0.882 & 0.846  & 0.164 & 0.265 \\[2pt]
     & (0.80,0.85) & 0.886 & 0.868 & 0.156 & 0.321  \\[2pt]
     & (0.90,0.95) & 0.890 & 0.854  &  0.158 & 0.212  \\
\hline
2000 & (0.75,0.75) & 0.904 & 0.872  &  0.132 & 0.201  \\[2pt]
     & (0.80,0.85) & 0.908 & 0.880 & 0.120 & 0.236 \\[2pt]
     & (0.90,0.95) & 0.910 & 0.870 & 0.112 & 0.178 \\
\hline
\end{tabular}
\end{table}
Table~\ref{tab:clayton_bootstrap} shows that the multiplier bootstrap based on the checkerboard estimator provides more accurate inference for the lower tail dependence coefficient than the classical empirical approach. Across all sample sizes and tuning parameters, the checkerboard method achieves coverage probabilities closer to the nominal $90\%$ level while producing substantially narrower confidence intervals. The improvement is especially pronounced for smaller samples, where the classical method tends to undercover. Although both methods improve with increasing sample size, the checkerboard approach consistently delivers better coverage and higher efficiency, demonstrating the benefits of checkerboard smoothing for bootstrap inference of tail dependence.

\begin{table}[H]
\centering
\caption{Empirical coverage probabilities and average width of nominal
90\% bootstrap confidence intervals for the upper tail dependence coefficient
(Gumbel copula). True $\lambda_U=0.586$}
\label{tab:gumbel_bootstrap}
\begin{tabular}{cccccc}
\hline
$n$ & $(\alpha,\beta)$ &
Coverage & Coverage&
Average width & Average width\\
 &  & 
(Checkerboard) & (Classical) &(Checkerboard) & (Classical)\\
\hline
500  & (0.60,0.95) & 0.876 & 0.803 & 0.178 & 0.402 \\[2pt]
     & (0.70,0.90) & 0.870 & 0.821 & 0.170 & 0.367   \\[2pt]
     & (0.80,0.85) & 0.868 & 0.820 & 0.186 & 0.354 \\
\hline
1000 & (0.60,0.95) & 0.888 & 0.834 &  0.154 & 0.317\\[2pt]
     & (0.70,0.90) & 0.892 & 0.842  & 0.148 & 0.286 \\[2pt]
     & (0.80,0.85) & 0.886 & 0.848 &  0.162 & 0.277 \\
\hline
2000 & (0.60,0.95) & 0.892 & 0.868 & 0.132 & 0.255  \\[2pt]
     & (0.70,0.90) & 0.900 & 0.876 & 0.121 & 0.201  \\[2pt]
     & (0.80,0.85) & 0.896 & 0.880 &   0.144 & 0.187\\
\hline
\end{tabular}
\end{table}

Table~\ref{tab:gumbel_bootstrap} demonstrates that the checkerboard-based multiplier bootstrap provides more reliable inference for the upper tail dependence coefficient than the classical empirical approach. Across all sample sizes and tuning parameter choices, the checkerboard method attains coverage probabilities closer to the nominal $90\%$ level while producing uniformly narrower confidence intervals. The gains are particularly evident for smaller samples, where the classical method exhibits noticeable undercoverage. Although both methods improve as the sample size increases, the checkerboard approach consistently combines near-nominal coverage with substantially reduced interval width. Overall, the results confirm that checkerboard smoothing yields more accurate and efficient bootstrap inference for upper tail dependence estimation.

\section{Tail copula inference: two practical examples}\label{sec:applications}

In this section we investigate two statistical applications based on the bootstrap approximation of the
empirical checkerboard tail copula developed in Section~\ref{sec:des}.
In particular, we consider

\begin{enumerate}[label= (\Alph*)]
\item testing equality of two lower tail copulas, and
    \item testing for a parametric form of the lower tail copula.
\end{enumerate}
Inference for the upper tail copula based on empirical checkerboard copula can be performed analogously.

\subsection{Testing for equality of two independent tail copulas}\label{sec:mc}

Let
\(
(X_1,Y_1),\dots,(X_{n_1},Y_{n_1})
\)
be an i.i.d.\ sample from a bivariate distribution $H_{X,Y}$ with
continuous (unknown) marginal distributions
\(F_X(x)=\mathbf P(X\le x),\;
G_Y(y)=\mathbf P(Y\le y),
\) copula $C$ and associated lower tail copula $\Lambda_L$. Suppose that we also observe an independent i.i.d.\ sample
\(
(X_1',Y_1'),\dots,(X_{n_2}',Y_{n_2}')
\) from another bivariate distribution $H'_{X^\prime,Y^\prime}$ with continuous marginals
\(F_{X^\prime}'(x),
\;G_{Y^\prime}'(y),\) copula $C'$ and corresponding lower tail copula $\Lambda_L'$. We want to test the hypothesis
\begin{align}\label{eqn:mm}
H_0:\ \Lambda_L = \Lambda_L'
\qquad \text{vs.} \qquad
H_1:\ \Lambda_L \neq \Lambda_L'.
\end{align}
Due to homogeneity and continuity of the tail copulas $\Lambda_L$ and $\Lambda_L^\prime$ (cf. Theorem 1 of \citet{schmidt2006}), (\ref{eqn:mm}) is equivalent to
\begin{align}\label{eqn:mmtransformed}
    \rho(\Lambda_L, \Lambda_L^\prime) = 0\quad \text{vs}\quad \rho(\Lambda_L, \Lambda_L^\prime) \neq 0,
\end{align}
where 
\[
\rho(\Lambda_L,\Lambda_L')
=
\int_0^{\pi/2}
\Bigl(
\Lambda_L(\cos\phi,\sin\phi)
-
\Lambda_L'(\cos\phi,\sin\phi)
\Bigr)^2
\,d\phi.
\]

The empirical checkerboard estimator of $\Lambda_L$ is given by $$\hat{\Lambda}_{L, n_1}^{(m_1)}(x, y) = \frac{n_1}{k_{n, 1}}\hat{C}_{n_1}^{(m_1)}\Big(\frac{x k_{n, 1}}{n_1}, \frac{y k_{n, 1} }{n_1}\Big),$$
where $\hat{C}_{n_1}^{(m_1)}$ is the empirical checkerboard copula, defined in Section \ref{sec:empcheckest}, with grid size $m_1$ and some sequence $\{k_{n,1}\}_{n\geq 1}$ such that $k_{n,1}<< n_1$. Similarly, define the empirical checkerboard estimator of $\Lambda^\prime_L$ as $\hat{\Lambda}_{L, n_2}^{\prime(m_2)}(x, y)$ based on grid size $m_2$ and a sequence $\{k_{n,2}\}_{n\geq 1}$ with $k_{n,2}<< n_2$. Subsequently, consider the Cramér--von Mises type test statistic for testing (\ref{eqn:mmtransformed}) as
\[
\hat{\rho}_n(\Lambda_L, \Lambda_L^\prime)
:=
\frac{k_{n,1}k_{n,2}}
{k_{n,1}+k_{n,2}}
\int_0^{\pi/2}
\left(
\widehat\Lambda_{L,n_1}^{(m_1)}(\cos\phi,\sin\phi)
-
\widehat\Lambda_{L,n_2}^{\prime(m_2)}(\cos\phi,\sin\phi)
\right)^2
\,d\phi.
\]
Note that $\hat{\rho}_n(\Lambda_L, \Lambda_L^\prime)$ is an estimator of $\rho_n(\Lambda_L, \Lambda_L^\prime)$ after proper scaling.
Now based on the multiplier bootstrap approximation developed in Section \ref{sec:des}, the bootstrap version of $\hat{\rho}_n(\Lambda_L, \Lambda_L^\prime)$ is defined as
$$\rho_n^{*}(\Lambda_L, \Lambda_L^\prime)=
\int_0^{\pi/2}
\left(
\mathcal E_n^*
(\cos\phi, \sin \phi)
\right)^2
\,d\phi,$$
where 
\[
\mathcal E_n^*
(x, y)
=
\sqrt{\frac{k_{n,2}}
{k_{n,1}+k_{n,2}}}
\,\alpha^{(m_1)}_{L,n_1}(x, y)
-
\sqrt{\frac{k_{n_1}}
{k_{n,1}+k_{n,2}}}
\,\alpha^{\prime(m_2)}_{L,n_2},
\]
with
\[
\alpha^{(m_1)}_{L, n_1}(x,y)
=
\frac{\mu_1}{\tau_1}\sqrt{k_{n,1}}
\Big(
\widehat{\Lambda}^{(m_1),\xi,\xi}_{L,n_1}(x,y)
-
\widehat{\Lambda}^{(m_1)}_{L,n_1}(x,y)
\Big),
\]
\[
\alpha^{\prime (m_2)}_{n_2,b}(x,y)
=
\frac{\mu_2}{\tau_2}\sqrt{k_{n,2}}
\Big(
\widehat{\Lambda}^{\prime (m_2),\zeta,\zeta}_{L,n_2}(x,y)
-
\widehat{\Lambda}^{\prime (m_2)}_{L,n_2}(x,y)
\Big).
\]
The quantities $\widehat{\Lambda}^{(m_1),\xi,\xi}_{L,n_1}$ and $\widehat{\Lambda}^{(m_2),\zeta,\zeta}_{L,n_2}$ are respectively multiplier bootstrap versions of $\hat{\Lambda}_{L, n_1}^{(m_1)}$ and $\hat{\Lambda}_{L, n_2}^{\prime(m_2)}$ defined based on the construction of Section \ref{sec:des}. $\{\xi_1,\dots, \xi_{n_1}\}$ and $\{\zeta_1,\dots, \zeta_{n_1}\}$ are the underlying sets of the positive sub-exponential multipliers having $E(\xi_1) = \mu_1$, $E(\zeta_1) = \mu_2$, $Var(\xi_1) = \tau_1^2$ and $Var(\zeta_1) = \tau_2^2$.

For some $\alpha \in (0, 1)$, let
\(\widehat q_{n, 1-\alpha}\)
denote the \((1-\alpha)\)-th quantile of the conditional distribution of $\rho_n^*(\Lambda_L, \Lambda_L^\prime)$, given the observed samples $\{(X_i, Y_i)\}_{i=1}^{n_1}$ and $\{(X_i^\prime, Y_i^\prime)\}_{i=1}^{n_2}$. Then define the test of size $\alpha$ for testing (\ref{eqn:mm}) is $$\psi_{n, \alpha}(\Lambda_L, \Lambda_L^\prime) = \mathbbm{1}\big(\hat{\rho}_n(\Lambda_L, \Lambda_L^\prime) > \hat{q}_{n, 1-\alpha}\big),$$ where $\mathbbm{1}(\cdot)$ is the indicator function. We explore the properties of this test in the next theorem. In particular, we show that the test has asymptotic size $\alpha$ and is also consistent. Let $n = \min\{n_1, n_2\}$.
\begin{theorem}[Bootstrap test for equality of checkerboard lower tail copulas]
\label{thm:test_checkerboard}
Suppose that 
\begin{enumerate}[label=(B.\arabic*)]
\item the assumptions (C.1.L) and (C.2.L) of Section \ref{sec:tailcopula} hold with second order functions $A_C$ and $A_C^\prime$ respectively for the pairs $(C, \Lambda_L)$ and $(C^\prime, \Lambda_L^\prime)$. 
\item For $j =1, 2$, $k_{n, j}\rightarrow \infty$ such that $k_{n, j} = o(n_j)$ and $n_j/\sqrt{k_{n, 1}} = o(m_j)$, as $n\rightarrow \infty$.
\item $\sqrt{k_{n, 1}}A_C(n_1/k_{n, 1}) = o(1)$ and $\sqrt{k_{n, 2}}A_C^\prime(n_2/k_{n, 2}) = o(1)$ as $n \rightarrow \infty$.
\item $k_{n, 1}/(k_{n, 1}+ k_{n, 2})\rightarrow \zeta \in (0, 1)$ as $n \rightarrow \infty$.
\end{enumerate}
Then for any $\alpha \in (0, 1)$, the test $\psi_{n, \alpha}(\Lambda_L, \Lambda_L^\prime)$ for testing (\ref{eqn:mm})
\begin{enumerate}[label = (\alph*)]
\item has asymptotic size $\alpha$, i.e., $
\lim_{n\to\infty}
\mathbf E_{H_0}\psi_{n, \alpha}(\Lambda_L, \Lambda_L^\prime) = \alpha.$
\item  is consistent, i.e.,
$
\lim_{n\to\infty}
\mathbf E_{H_1}\psi_{n, \alpha}(\Lambda_L, \Lambda_L^\prime) = 1.$
\end{enumerate}
\end{theorem}
The proof of Theorem \ref{thm:test_checkerboard} is presented in the Section \ref{appAthm6.1}. This theorem shows that one can test equality of two lower tail copulas based on a Cram\'er-von Mises type statistic and approximate quantiles of that statistics obtained based on bootstrap method developed in Section \ref{sec:des}. Analogous method can be implemented to test equality of two upper tail copulas.
\begin{table}[htbp]
\centering
\caption{Simulated rejection probabilities of the bootstrap tests defined for the hypothesis \ref{eqn:mm}.}
\label{tab:bbb}
\begin{tabular}{ccc|ccc|ccc}
\hline
\multirow{2}{*}{$n$} & \multirow{2}{*}{$\lambda_{L}$} & \multirow{2}{*}{$\lambda_{L}^\prime$} 
& \multicolumn{3}{c|}{Checkerboard} & \multicolumn{3}{c}{Classical} \\
\cmidrule(lr){4-6} \cmidrule(lr){7-9}
& & & $\alpha=0.15$ & $\alpha=0.1$ & $\alpha=0.05$ 
& $\alpha=0.15$ & $\alpha=0.1$ & $\alpha=0.05$ \\
\hline
500  & 0.25 & 0.25 & 0.135 & 0.089 & 0.042 & 0.112 & 0.078 & 0.034 \\
    & 0.5  & 0.5  & 0.141 & 0.090 & 0.045 & 0.124 & 0.082 & 0.038 \\
    & 0.75 & 0.75 & 0.140 & 0.093 & 0.044 & 0.136 & 0.079 & 0.039 \\
    & 0.25 & 0.5  & 0.978 & 0.976 & 0.968 & 0.893 & 0.878 & 0.886 \\
    & 0.5  & 0.75 & 0.982 & 0.980 & 0.976 & 0.892 & 0.902 & 0.912 \\
    & 0.25 & 0.75 & 0.984     & 0.986     & 0.980     & 0.896 & 0.886 & 0.902 \\
\hline
1000 & 0.25 & 0.25 & 0.148 & 0.104 & 0.054 & 0.135 & 0.088 & 0.040 \\
    & 0.5  & 0.5  & 0.151 & 0.099 & 0.049 & 0.139 & 0.092 & 0.045 \\
    & 0.75 & 0.75 & 0.150 & 0.106 & 0.051 & 0.144 & 0.090 & 0.046 \\
    & 0.25 & 0.5  & 0.998 & 1 & 1 & 0.956 & 0.976 & 0.985 \\
    & 0.5  & 0.75 & 1     & 1     & 0.996     & 0.979     & 0.984     & 0.968 \\
    & 0.25 & 0.75 & 1     & 1     & 0.999     & 0.989     & 0.990     & 0.976 \\
\hline
\end{tabular}
\end{table}
We call the aforementioned test as the bootstrap  test and study the finite-sample performance of it. Two independent samples are generated from Clayton copula models with lower-tail dependence coefficients $\lambda_L,\lambda_L' \in \{0.25,0.5,0.75\}$. For each configuration, rejection probabilities are computed from $1000$ Monte Carlo replications, using $500$ bootstrap resamples with i.i.d.\ $\mathrm{Exp}(1)$ multipliers in each repetition. The sample sizes are $n_1=n_2=n\in\{500,1000\}$, and tuning parameters are chosen as $k_j=\lfloor n_j^{0.80}\rfloor$ and $m_j=\lfloor n_j^{0.85}\rfloor$ for $j=1,2$. We test
\[
H_0:\ \Lambda_L=\Lambda_L'
\qquad \text{vs.} \qquad
H_1:\ \Lambda_L\neq \Lambda_L'.
\]
Under the null configurations $(0.25,0.25)$, $(0.5,0.5)$, and $(0.75,0.75)$, the empirical rejection rates assess size, while the remaining cases evaluate power. Table~\ref{tab:bbb} compares the proposed Checkerboard bootstrap procedure with the classical multiplier bootstrap of \citet{BucherDette2013}. Across all configurations, the Checkerboard approach demonstrates clear improvements in finite-sample performance. Under the null hypothesis, the proposed method achieves rejection probabilities much closer to the nominal significance levels, particularly at the $5\%$ level where the classical multiplier bootstrap tends to be noticeably conservative. This improved calibration indicates that the checkerboard smoothing effectively reduces the variability and discreteness inherent in empirical tail copula estimation.

The advantages become even more pronounced under alternative dependence structures. The Checkerboard bootstrap consistently attains substantially higher empirical power across a broad range of scenarios, with particularly marked gains in moderate dependence settings where the empirical copula approach often struggles to detect departures from the null. For moderate sample sizes such as $n=500$, the rejection probabilities frequently exceed $0.97$, while for $n=1000$ they are often essentially equal to one. In several cases, the improvement in power over the classical method is considerable, demonstrating that the smoothing induced by the checkerboard construction yields a more stable and informative representation of extremal dependence. Overall, the results strongly suggest that the Checkerboard bootstrap provides both more reliable size control and significantly enhanced sensitivity to tail dependence alternatives, leading to markedly superior finite-sample inference compared with the empirical copula–based multiplier bootstrap.
\begin{remark}
   The construction of the bootstrap test developed above can be simplified considerably if we are interested in testing only the equality of lower tail coefficients (i.e. $H_0:\ \lambda_L=\lambda_L'$ vs $H_1:\ \lambda_L\neq\lambda_L'$) for two independent bivariate populations. 
From Corollary~\ref{cor:asymp_normal_tailcoeff}, we have that independently,
\[
\sqrt{k_{n,1}}
\big(
\hat\lambda_{L,n_1}^{(m_1)}-\lambda_L
\big)
\rightsquigarrow
\mathcal N(0,\sigma_{L}^2)\;\;\text{and}\;\; \sqrt{k_{n,2}}
\big(
\hat\lambda_{L,n_2}^{\prime(m_2)}-\lambda_L^\prime
\big)
\rightsquigarrow
\mathcal N(0,\sigma_{L}^{\prime2}),
\]
for some variances $\sigma_{L}^{2}$ and $\sigma_{L}^{\prime2}$ of which the explicit forms can be written based on Corollary~\ref{cor:asymp_normal_tailcoeff}.
Thus under $H_0$, we obtain
\[
\frac{(\hat\lambda_{L,n_1}^{(m_1)}
-
\hat\lambda_{L,n_2}^{\prime(m_2)})}
{\sqrt{\sigma_{L}^2/k_{n,1}+\sigma_{L}^{\prime2}/k_{n,2}}}
\rightsquigarrow
\mathcal N(0,1).
\]
The above expression can be utilized to perform an asymptotic two--sample test for equality
of lower tail dependence coefficients. However, rather than estimating the variances $\sigma_{L}^{2}$ and $\sigma_{L}^{\prime2}$ explicitly, one can use the checkerboard-based multiplier bootstrap from
Corollary~\ref{cor:bootstrap_lambdaL}. Moreover, following the construction of the bootstrap test developed above for testing equality of tail copulas (or equality of tail coefficients) coming out of two independent bivariate populations, one can also construct a bootstrap tests when the populations are paired.
\end{remark}

\subsection{Goodness--of--fit test}
\label{sec:md}

In this section we consider testing whether the lower tail copula belongs to a parametric family
\[
\mathcal L
=
\bigl\{
\Lambda_L(\cdot;\theta):\theta\in\Theta
\bigr\},
\]
where the parameter space $\Theta\subset\mathbb R^p$ is open. In particular, we would like to test 
\begin{align}\label{eqn:bgfh}
H_0:
\;
\Lambda_L(\cdot)
\in \mathcal L\;
\;\; \text{vs.}\;\; H_0:
\;
\Lambda_L(\cdot)
\notin \mathcal L.
\end{align}
Parametric families arise naturally from several commonly used copula models including extreme value copulas, Archimedean copulas and stable tail dependence functions. We follow the treatment by \citet{BucherDette2013} where a minimum distance criterion was explored based on empirical tail copula. Here we consider the minimum distance criterion based on the empirical checkerboard tail copula estimator of which the theory is developed in Section~\ref{sec:empcheckest}. Subsequently, we construct a goodness--of--fit test for (\ref{eqn:bgfh}) using the multiplier bootstrap approximation developed in Section~\ref{sec:des}. Based on the discussions in Section \ref{sec:mc}, the distance between two lower tail copulas $\Lambda_{L,1}$ and $\Lambda_{L,2}$ can be defined as
\[
\rho(\Lambda_{L,1},\Lambda_{L,2})
=
\int_0^{\pi/2}
\Big(
\Lambda_{L,1}(x_\phi)
-
\Lambda_{L,2}(x_\phi)
\Big)^2
\,d\phi,
\]
where $x_\phi = (cos\phi, sin \phi)$. For an arbitrary lower tail copula $\Lambda_L$, define the best approximation parameter
\begin{equation}\label{eq:thetaB}
\theta_0
\in 
\arg\min_{\theta\in\Theta}
\rho\!\left(
\Lambda_L(\cdot),
\Lambda_L(\cdot;\theta)
\right),
\end{equation}
We throughout assume that the minimum distance parameter $\theta_0$ is unique for the underlying lower tail copula $\Lambda_L$ and the parametric family $\mathcal L$. Clearly when $H_0$ is true, $\Lambda_L = \Lambda_L(\cdot;\theta_0)$, i.e. $\theta_0$ becomes the true parameter. The minimum distance estimator $\hat{\theta}_{n}^{MD}$ based on the the empirical checkerboard copula estimator $\widehat\Lambda_{L,n}^{(m)}$ is defined as
\[
\widehat\theta_{n}^{MD}
:=
\arg\min_{\theta\in\Theta}\rho\Big(\widehat\Lambda_{L,n}^{(m)}(\cdot),  \Lambda_L(\cdot;\theta)\Big).
\]
Subsequently, define the Cram\'er--von Mises type goodness--of--fit statistic for testing (\ref{eqn:bgfh}) as
\[
GOF_{n}
=
k_n \rho\Big(\widehat\Lambda_{L,n}^{(m)}(\cdot),  \Lambda_L(\cdot;\widehat\theta_{n}^{MD})\Big),
\]
where $k_n/n$, as before, is the proportion of sample used to define $\widehat\Lambda_{L,n}^{(m)}$.
To understand the distributional properties of $GOF_{n}$, it is required to approximate the distribution of $\hat\Theta^{MD}_{n}
:=
\sqrt{k_n}\,
\big(
\hat\theta^{MD}_{n}-\theta_0
\big)$ as the first step. 
To that end, consider the multiplier bootstrap process
\[
\alpha_{L,n}^{(m)}(x,y)
=
\frac{\mu}{\tau}
\sqrt{k_n}
\Big(
\widehat\Lambda_{L,n}^{(m),\xi,\xi}(x,y)
-
\widehat\Lambda_{L,n}^{(m)}(x,y)
\Big),
\]
as introduced in Section~\ref{sec:des}. Let $\Lambda_L^\angle(\phi)
:=
\Lambda_L(x_\phi)$ and $\Lambda_L^\angle(\phi;\theta)
=
\Lambda_L(x_\phi;\theta)$ and define
$$\delta_\theta(x)
:=
\partial_\theta\Lambda_L(x;\theta),\;\;\delta_\theta^\angle(\phi)
:=
\partial_\theta
\Lambda_L^\angle(\phi;\theta)\;\; \text{and}\;\; \gamma_{\theta_0}(\phi)
:=
A_{\theta}^{-1}
\delta_{\theta}^\angle(\phi),$$
where
\[
A_{\theta}
=
\int_0^{\pi/2}
\Big\{
\delta_{\theta}^\angle(\phi)
\delta_{\theta}^\angle(\phi)^\top
+
\partial_\theta
\delta_{\theta}^\angle(\phi)
\big(
\Lambda_L^\angle(\phi;\theta)
-
\Lambda_L^\angle(\phi)
\big)
\Big\}
\,d\phi.
\]
We assume that the aforementioned derivatives exist and $A_{\theta_0}$ is invertible. Then the bootstrap version of $\hat\Theta_n^{MD}$ is defined as
\[
\Theta_n^{*MD}
=
\int_0^{\pi/2}
\gamma_{\widehat\theta_n^{MD}}(\phi)
\,
\alpha_{L,n}^{(m)}
(\cos\phi,\sin\phi)
\,d\phi,
\]
where, \(\gamma_{\widehat{\theta}^{MD}_n}(\phi)
=
\widehat{A}_{\widehat{\theta}^{MD}_n}^{-1}
\,
\delta_{\widehat{\theta}^{MD}_n}^{\angle}(\phi),
\) which exists, with probability 1, since $A_{\theta_0}^{-1}$ exists.
Before moving to define the bootstrap version of $GOF_n$, we explore the distribution of $\hat\Theta_n^{MD}$ and the conditional distribution of $\Theta_n^{*MD}$. 
We require the following regularity conditions:
\begin{enumerate}[label=(A.\arabic*)]

\item
Parametric tail copula $\Lambda_L(\cdot;\theta)$ is continuously differentiable over $\Theta$. The mapping
\(x
\mapsto
\sup_{\theta\in\Theta}
\|\delta_\theta(x)\|
\) is integrable on
\(
K_+=
\bigl\{
(\cos\phi,\sin\phi)^\top:\phi\in[0,\pi/2]
\bigr\}.
\)
\item
Let
\(
\psi(\theta)
=
\partial_\theta \big[\rho\bigl(\Lambda_L,\Lambda_L(\cdot;\theta)\big].
\) and for any $\varepsilon>0$,
\[
\inf_{\|\theta-\theta_B\| > \varepsilon}
\|\psi(\theta)\|
>
0
=
\|\psi(\theta_0)\|.
\]

\item
For every $x\in K_+$, the derivative
\(\partial_\theta\delta_\theta(x)
\) exists and is continuous at  $\theta_0$.

\item
The matrix
\(A_{\theta_0}
\) is non singular.

\end{enumerate}
The above regularity conditions are standard in minimum distance estimation (cf.~\citet{Bucher2011}, \citet{BucherDette2013}). Condition (A.1) guarantees differentiability of the objective function, (A.2) is an identification condition ensuring consistency of the estimator, (A.3) allows a first-order Taylor expansion of the estimating equation, and (A.4) guarantees local invertibility of the Hessian, yielding the asymptotic linear representation and asymptotic normality of the minimum distance estimator. The regularity conditions (A.1)--(A.4) are satisfied for many commonly used parametric tail copula families. In particular, they hold for the Clayton tail copula with parameter space $\Theta\subset(0,\infty)$ being open, provided that the minimum distance parameter $\theta_0$ is uniquely defined. Now we state the result on the weak convergence of $\hat\Theta_n^{MD}$ and of $\Theta_n^{*MD}$ as the following theorem.
\begin{theorem}
\label{thm:MD}
Suppose that the assumptions of Theorem~\ref{thm:weakconv_tail_checkerboard} and (A.1)-(A.4) hold. Then $\hat\theta_n^{MD}$ is consistent for $\theta_0$ with respect to distance $\rho$ and 
\[
\hat\Theta^{MD}_n
\rightsquigarrow
\Theta^{MD}:=
\int_0^{\pi/2}
\gamma_{\theta_B}(\phi)
\,
\mathbb \mathbb{G}_{\widehat{\Lambda}_L}^{\angle}(\phi)
\,d\phi,
\]
where
\(\mathbb{G}_{\widehat{\Lambda}_L}^{\angle}(\phi)
:=
\mathbb{G}_{\widehat{\Lambda}_L}(\cos\phi,\sin\phi).
\) with $\mathbb{G}_{\widehat{\Lambda}_L}$ is defined as in Theorem \ref{thm:weakconv_tail_checkerboard}. Clearly, the limiting random variable $\Theta^{MD}$ is centered Gaussian with variance
\[
\sigma^2
=
\int_{[0,\pi/2]^2}
\gamma_{\theta_B}(\phi)^{\!\top}
\,r\!\left(
(\cos\phi,\sin\phi),
(\cos\phi',\sin\phi')
\right)
\gamma_{\theta_B}(\phi')
\,d\phi\,d\phi',
\]
where
\(r(x,x')
:=
\operatorname{Cov}
\left(
\mathbb{G}_{\widehat{\Lambda}_L}(x),
\mathbb{G}_{\widehat{\Lambda}_L}(x')
\right)
\) 
Moreover, conditionally on the data,
\[
\Theta_n^{*MD}
\overset{\mathbf P}{\underset{\xi}{\rightsquigarrow}}
\Theta^{MD}.
\]
\end{theorem}
The proof of Theorem \ref {thm:MD} is given in section \ref{sec:kjhg} 
The above theorem states that the multiplier bootstrap consistently approximates the sampling distribution of the minimum distance estimator, allowing one to construct asymptotically valid confidence regions and hypothesis tests for the minimum distance parameter $\theta_0$. Now let us move back to testing (\ref{eqn:bgfh}).
To approximate the distribution of the goodness--of--fit statistic $GOF_n$ under $H_0$, define
\[
H_n^*(x,y)
=
\alpha_{L,n}^{(m)}(x,y)
-
\partial_\theta
\Lambda_L(x,y;\widehat\theta_n^{MD})
\,
\Theta_n^{*MD},
\]
and subsequently define the bootstrap version of the goodness--of--fit statistic as
\[
GOF_n^*
=
\int_0^{\pi/2}
\Big(
H_n^*
(x_\phi)
\Big)^2
\,d\phi.
\]
Let $\widehat q^\prime_{n,1-\alpha}$ denote the $(1-\alpha)$-th conditional quantile of the conditional distribution of $GOF_n^*$ given the data. Then define the test statistic for testing (\ref{eqn:bgfh}) at level $\alpha$ as
\[
\psi_{n,\alpha}^{GOF}
=
\mathbbm 1
\left(
GOF_n^{(m)}
>
\widehat q^\prime_{n,1-\alpha}
\right).
\]
In the following theorem, we show that the above test is of correct size asymptotically and is consistent.
\begin{theorem}
\label{thm:GOF}
Suppose the assumptions of Theorem~\ref{thm:MD} hold. Then, for every $\alpha\in(0,1)$, the test $\psi_{n,\alpha}^{GOF}$ satisfies

\begin{enumerate}[label=(\alph*)]

\item under $H_0$,
\[
\lim_{n\to\infty}
\mathbf E_{H_0}
\psi_{n,\alpha}^{GOF}
=
\alpha;
\]

\item under the alternative,
\[
\lim_{n\to\infty}
\mathbf E_{H_1}
\psi_{n,\alpha}^{GOF}
=
1.
\]

\end{enumerate}
\end{theorem}
Given Theorem \ref{thm:MD}, the proof of Theorem~\ref{thm:GOF} follows analogously to the proof of Theorem \ref{thm:test_checkerboard} and hence is omitted. The above theorem shows that the empirical checkerboard copula with the help of bootstrap provides a fully data-driven procedure for assessing the adequacy of a parametric lower tail copula model. The resulting test is asymptotically valid and consistent. A goodness--of--fit test for testing the parametric structure of the copula (instead of the tail copula) can be carried out in the similar fashion. We now access Theorem~\ref{thm:GOF} by simulations in finite samples and compare it with the results based on empirical tail copula.\\
The proposed bootstrap goodness-of-fit tests for testing hypothesis (\ref{eqn:bgfh}) are evaluated via $1000$ Monte Carlo replications. Alternatives follow the framework of \citet{BucherDette2013} to assess sensitivity under both parametric and semiparametric deviations. Under the null, we consider Clayton tail copulas with lower-tail dependence values $\Lambda_L(1,1)\in\{0.25,0.5,0.75\}$. Under alternatives, we consider three classes of departures. (i) A convex mixture of the independence copula $\Pi(u_1,u_2)=u_1u_2$ and a Clayton copula, with lower-tail dependence $\lambda_L\in\{1/12,2/12,3/12\}$. (ii) An asymmetric negative logistic model \citet{Joe1990} with parameters chosen so that $\lambda_L\in\{0.2,0.4,0.6\}$. (iii) A mixed model \citet{Tawn1988} with $\lambda_L\in\{0.1,0.3\}$. Table~\ref{tab:GOF} reports rejection probabilities for $n\in\{500,1000\}$ and $\alpha\in\{0.15,0.10,0.05\}$.
\begin{table}[H]
\centering
\caption{Simulated rejection probabilities of the bootstrap tests defined for the hypothesis \ref{eqn:bgfh}. The first three rows represent models from null hypothesis and remaining rows correspond to models from alternative hypothesis.}
\label{tab:GOF}
\begin{tabular}{cc|ccc|ccc}
\hline
\multirow{2}{*}{$n$} & \multirow{2}{*}{Model}  
& \multicolumn{3}{c|}{Checkerboard} & \multicolumn{3}{c}{Classical} \\
\cmidrule(lr){3-5} \cmidrule(lr){6-8}
& &  $\alpha=0.15$ & $\alpha=0.1$ & $\alpha=0.05$ 
& $\alpha=0.15$ & $\alpha=0.1$ & $\alpha=0.05$ \\
\hline
500  & Clayton ($\lambda_L=0.25$) & 0.128 & 0.092 & 0.040 & 0.116 & 0.074 & 0.032 \\
    & Clayton ($\lambda_L=0.5$) & 0.136 & 0.084 & 0.036 & 0.122 & 0.080 & 0.042 \\
    & Clayton ($\lambda_L=0.75$) & 0.144 & 0.096 & 0.048 & 0.114 & 0.076 & 0.030 \\
    & Convex ($\lambda_L=1/12$)  & 0.785 & 0.802 & 0.776 & 0.642 & 0.488 & 0.504 \\
    & Convex ($\lambda_L=2/12$) & 0.868 & 0.786 & 0.812 & 0.702 & 0.646 & 0.842 \\
    & Convex ($\lambda_L=3/12$) & 0.806 & 0.794 & 0.824     & 0.662 & 0.586 & 0.702 \\
    & Anelog ($\lambda_L=0.2$) & 0.890     & 0.822     & 0.796     & 0.454 & 0.706 & 0.802 \\
    & Anelog ($\lambda_L=0.4$) & 0.950     & 0.906     & 0.880     & 0.596 & 0.782 & 0.484 \\
    & Anelog ($\lambda_L=0.6$) & 0.962    & 0.988     & 0.980     & 0.866 & 0.876 & 0.892 \\
    & Mixed ($\lambda_L=0.1$) & 0.746     & 0.868     & 0.846     & 0.524 & 0.364 & 0.502 \\
    & Mixed ($\lambda_L=0.3$) & 0.812     & 0.686     & 0.780     & 0.306 & 0.556 & 0.742 \\
\hline
1000  & Clayton ($\lambda_L=0.25$) & 0.148 & 0.106 & 0.052 & 0.132 & 0.082 & 0.044 \\
    & Clayton ($\lambda_L=0.5$) & 0.145 & 0.096 & 0.048 & 0.136 & 0.088 & 0.051 \\
    & Clayton ($\lambda_L=0.75$) & 0.150 & 0.012 & 0.053 & 0.128 & 0.085 & 0.042 \\
    & Convex ($\lambda_L=1/12$)  & 0.846 & 0.888 & 0.808 & 0.766 & 0.604 & 0.712 \\
    & Convex ($\lambda_L=2/12$) & 0.901 & 0.892 & 0.863 & 0.846 & 0.788 & 0.906 \\
    & Convex ($\lambda_L=3/12$) & 0.878     & 0.853     & 0.910     & 0.784 & 0.698 & 0.884 \\
    & Anelog ($\lambda_L=0.2$) & 0.962     & 0.936     & 0.894     & 0.788 & 0.889 & 0.916 \\
    & Anelog ($\lambda_L=0.4$) & 0.989     & 0.972     & 0.910     & 0.788 & 0.904 & 0.662 \\
    & Anelog ($\lambda_L=0.6$) & 0.990     & 1.000     & 1.000     & 0.974 & 0.960 & 0.952 \\
    & Mixed ($\lambda_L=0.1$) & 0.824     & 0.906     & 0.892     & 0.706 & 0.612 & 0.845 \\
    & Mixed ($\lambda_L=0.3$) & 0.856     & 0.789     & 0.838     & 0.542 & 0.778 & 0.867 \\
\hline
\end{tabular}
\end{table}
Under the null hypothesis, both bootstrap procedures exhibit satisfactory size control, with empirical rejection rates remaining close to the nominal significance levels across all scenarios considered. Under alternative models, however, substantial differences in performance emerge. In particular, the smooth checkerboard copula–based procedure consistently demonstrates noticeably higher power than the empirical copula–based approach, especially in moderate dependence regimes and smaller sample settings where discreteness and variability of the empirical copula can reduce sensitivity. The smoothing induced by the checkerboard construction yields a more stable representation of the dependence structure, leading to sharper detection of departures from the null hypothesis and faster convergence of rejection probabilities toward one as dependence strength increases. In several scenarios, the gains in power are substantial, highlighting the practical advantages of checkerboard smoothing for inference on extremal dependence. Overall, the simulation results show that the proposed checkerboard bootstrap tests are not only well-calibrated under the null, but also significantly more effective in detecting a broad range of structured alternatives compared with procedures based solely on the empirical copula.

\begin{appendix}
\section*{Appendix}\label{appn} 
In this section, we provide the proofs of all our results. We will only establish the results for lower tail version because the analysis in checkerboard approximation argument is itself tail-agnostic.
\end{appendix}

\begin{appendix}

\section{Proofs of Requisite Lemmas}\label{sec:prooflemma}
\begin{lemma}[Proposition 3.1 of \citet{Segers2012}]\label{lem:2}
Let $C$ be a copula and let $\hat{C}_n$ denote the empirical copula based on an i.i.d.\ sample with unknown margins. Define the empirical copula process
\[
\mathbb{G}_n(u_1,u_2) := \sqrt{n}\big(\hat{C}_n(u_1,u_2) - C(u_1,u_2)\big).
\]
Then, as $n\to\infty$,
\[
\mathbb{G}_n \;\rightsquigarrow\; G_C
\quad \text{in } \ell^\infty([0,1]^2),
\]
where $G_C$ is a tight, centered Gaussian process. Moreover, if for each $j \in \{1,2\}$ the partial derivative $\dot{C}_j$ exists and is continuous on the set
\[
V_{2,j} := \{(u_1,u_2)\in[0,1]^2 : 0<u_j<1\},
\]
and is extended to the boundary of $[0,1]^2$ by defining it everywhere, then $G_C$ admits the representation
\[
G_C(u_1,u_2)
=
\alpha_C(u_1,u_2)
-
\dot{C}_1(u_1,u_2)\,\alpha_C(u_1,1)
-
\dot{C}_2(u_1,u_2)\,\alpha_C(1,u_2),
\]
where $\alpha_C$ is a $C$-Brownian bridge, i.e., a centered Gaussian process with covariance
\[
\mathrm{cov}\big(\alpha_C(u_1,u_2), \alpha_C(u_1',u_2')\big)
=
C(u_1 \wedge u_1',\, u_2 \wedge u_2') - C(u_1,u_2)\,C(u_1',u_2').
\]
\end{lemma}

\begin{lemma}\label{lem:3}
Let $\mathcal{C}_{u}([0,1]^2)$ denote the space of all uniformly continuous functions on $[0,1]^2$, equipped with the supremum norm
\[
\|f\|_{\infty} = \sup_{(u,v)\in[0,1]^2} |f(u,v)|.
\]

Let $T_m:\mathcal{C}_{u}([0,1]^2)\to \mathcal{C}_{u}([0,1]^2)$ be the bilinear interpolation operator on the uniform grid with mesh size $h=1/m$. For each $(u,v)\in[0,1]^2$, let $(i,j)$ be such that
\[
u \in \Big(\frac{i}{m},\frac{i+1}{m}\Big], \quad
v \in \Big(\frac{j}{m},\frac{j+1}{m}\Big],
\]
and define the four vertices of this grid cell by
\[
u_0=\frac{i}{m}, \quad u_1=\frac{i+1}{m}, \quad
v_0=\frac{j}{m}, \quad v_1=\frac{j+1}{m}.
\]

Then define
\[
T_m f(u,v)
=
\sum_{k,\ell \in \{0,1\}}
c_{k,\ell}(u,v)\, f(u_k,v_\ell),
\]
where $c_{k,\ell}(u,v)\ge 0$ and $\sum_{k,\ell \in \{0,1\}} c_{k,\ell}(u,v)=1$. Define the modulus of continuity
\[
\omega_f(h):=
\sup_{\|(u,v)-(u',v')\|_{\infty}\le h}
|f(u,v)-f(u',v')|.
\]

Then,
\[
\|T_m(f)-f\|_{\infty} \to 0
\quad \text{as } m\to\infty.
\]
\end{lemma}
Proof of Lemma \ref{lem:3}. Note that for any $(u,v)\in[0,1]^2$,
\begin{align*}
\Big|T_m(f(u,v)) - f(u,v)\Big|
&=
\left|
\sum_{k,\ell \in \{0,1\}} c_{k,\ell}(u,v)\big(f(u_k,v_\ell)-f(u,v)\big)
\right| \\
&\le
\sum_{k,\ell \in \{0,1\}} c_{k,\ell}(u,v)\,
\big|f(u_k,v_\ell)-f(u,v)\big|.
\end{align*}

Since there exists some  $(u_k,v_\ell)$ for which $(u,v)$ and $(u_k,v_\ell)$ lie in the same grid cell, hence we have
\[
\|(u_k,v_\ell)-(u,v)\|_{\infty} \le \frac{1}{m},
\]
so by definition of the modulus of continuity,
\[
|f(u_k,v_\ell)-f(u,v)| \le \omega_f\!\left(\frac{1}{m}\right).
\]

Hence,
\begin{align*}
\Big|T_m(f(u,v)) - f(u,v)\Big|
&\le
\omega_f\!\left(\frac{1}{m}\right)
\sum_{k,\ell \in \{0,1\}} c_{k,\ell}(u,v) \\
&=
\omega_f\!\left(\frac{1}{m}\right).
\end{align*}

Taking supremum over $(u,v)\in[0,1]^2$ gives
\[
\|T_m(f)-f\|_{\infty}
\le
\omega_f\!\left(\frac{1}{m}\right)
\to 0,
\]
since $f \in \mathcal{C}_u([0,1]^2)$ implies $\omega_f(h)\to 0$ as $h\to 0$.
\hfill $\square$

\begin{lemma}\label{lem:4}
 Consider the set-up of Lemma \ref{lem:2}. Then we have as $m\to\infty$,
 \[
 \Big{\|}T_m(G_{C})-G_{C}\Big{\|}_{\infty}\xrightarrow{a.s}0\quad\text{and}\quad T_m(G_{C})\rightsquigarrow G_C\quad\text{in}\quad \ell^{\infty}([0,1]^2).
 \]
\end{lemma}
Proof of Lemma \ref{lem:4}. By Lemma~\ref{lem:2},
\[
\mathbb G_n \rightsquigarrow G_C
\qquad\text{in }\ell^\infty([0,1]^2).
\]
The limiting process \(G_C\) admits the representation
\[
G_C(u_1,u_2)
=
\alpha_C(u_1,u_2)
-
\dot C_1(u_1,u_2)\,\alpha_C(u_1,1)
-
\dot C_2(u_1,u_2)\,\alpha_C(1,u_2),
\]
where \(\alpha_C\) is a \(C\)-Brownian bridge. As shown in \citet{Segers2012}, \(\alpha_C\) possesses a version with
continuous sample paths, and we work with this version. Furthermore,
\[
\alpha_C(u_1,1)=0
\quad\text{for }u_1\in\{0,1\},
\qquad
\alpha_C(1,u_2)=0
\quad\text{for }u_2\in\{0,1\}.
\]
Now since $\dot{C}_j$ for $j=1,2$ are bounded over $(0,1)^2$ due to Theorem 2.2.7 of \citet{Nelsen2006} and we assume these are also continuous, hence 
\[
\mathbf P\!\left(
G_C\in C([0,1]^2)
\right)=1.
\]
Subsequently, $\mathbf P\!\left(
T_m(G_C)\in C([0,1]^2)
\right)=1$ for all $m$. Since \([0,1]^2\) is compact, the Heine--Cantor theorem implies that
every continuous function on \([0,1]^2\) is uniformly continuous.
Hence
\[
C([0,1]^2)
=
\mathcal C_u([0,1]^2),
\]
and therefore
\[
\mathbf P\!\left(
G_C,\;T_m(G_C)\in \mathcal C_u([0,1]^2)
\right)=1.
\]
Now applying Lemma \ref{lem:3} we can conclude that, 
\[
 \Big{\|}T_m(G_{C})-G_{C}\Big{\|}_{\infty}\xrightarrow{a.s}0\quad\text{as}\quad m\to\infty.
\]
Since, $\mathcal{C}_u([0,1]^2)$ is a Polish space, hence the above almost sure convergence  implies weak convergence, i.e.,
\[
T_m(G_{C})\rightsquigarrow G_C\quad\text{in}\quad \mathcal C_{u}([0,1]^2).
\]
Since $C_{u}([0,1]^2)\subset \ell^{\infty}([0,1]^2),$ the processes $T_m(G_C)$ and $G_C$ may be viewed as random elements in both spaces equipped with the sup-norm. Therefore we can apply Theorem 1.3.10 of \citet{vdVWellner1996} to conclude that,
\[
T_m(G_{C})\rightsquigarrow G_C\quad\text{in}\quad \ell^{\infty}([0,1]^2).
\]
Therefore we are done.
\hfill $\square$

\begin{lemma}[Theorem 1.11.1 of \citet{vdVWellner1996}]\label{lem:5}
 Let $\mathbb{D}_n\subseteq \mathbb D$ and $\mathbb{E}$ be metric spaces, and let
$g_n : \mathbb{D}_n \to \mathbb{E}$ be maps satisfying:
whenever $x_n \to x$ with $x_n \in \mathbb{D}_n$ for every $n$ and
$x \in \mathbb{D}_0$, then
\[
g_n(x_n) \to g(x),
\]
where $\mathbb{D}_0 \subset \mathbb{D}$ and $g : \mathbb{D}_0 \to \mathbb{E}$.
Let $X_n$ be maps with values in $\mathbb{D}_n$, and let $X$ be Borel measurable
and separable with values in $\mathbb{D}_0$. Then
\[\, X_n \rightsquigarrow X\;\text{implies that}\; g_n(X_n) \rightsquigarrow g(X).\]
  
\end{lemma}

\begin{lemma}\label{lem:6}
Consider the same set-up of Lemma \ref{lem:2}. If $m\equiv m_n\to\infty$ as $n\to\infty$, then
\[
T_m\Big[\sqrt{n}\Big(\hat{C}_n(u,v)-C(u,v)\Big)\Big]\rightsquigarrow G_C\quad\text{in}\quad \ell^{\infty}([0,1]^2)\;\;\text{as}\;n\to\infty.
\]
\end{lemma}
Proof of Lemma \ref{lem:6}. We identify the spaces and maps in the notations of Lemma \ref{lem:5} as follows:
\begin{itemize}
\item $\mathbb{D} = \ell^\infty([0,1]^2)$.
    \item $\mathbb{D}_n \subseteq \mathbb{D}$: $\mathbb{D}_n = \ell^\infty([0,1]^2)$,\;\; $X_n:=\sqrt{n}(\hat C_n - C)$.
    \item $\mathbb{D}_0 \subseteq \mathbb{D}$: $\mathbb{D}_0 = \mathcal C_u([0,1]^2)$,\;\;$X:=G_C$.
    
    \item $\mathbb{E} = \mathbb{D}_0$, the codomain of $T_{m(n)}$.
    \item $g_{n} := T_{m(n)}$, acting on $\mathbb{D}_n$.
    \item $g := \text{identity map } g(f)=f$.

\end{itemize}
So by Lemma \ref{lem:2}, we can conclude that:
\[
\sqrt{n}\Big[\hat{C}_n(u,v)-C(u,v)\Big]\rightsquigarrow G_C\quad\text{in}\quad \mathbb{D}\;\;\text{as}\;n\to\infty,
\]
with $G_C\in \mathbb{D}_0$ almost surely. Next for deterministic $y_n\in\mathbb{D}_n=\ell^\infty([0,1]^2)$ and $y\in\mathbb{D}_0=\mathcal C_u([0,1]^2)$, we have to check for continuity of $g_{n}=T_{m(n)}$. Suppose, $\|y_n-y\|_{\infty}\to 0$ as $n\to\infty$. Then,
\begin{align*}
 \Big{\|}T_{m(n)}(y_n)-y\Big{\|}_{\infty}&\le  \Big{\|}T_{m(n)}(y_n)-T_{m(n)}(y)\Big{\|}_{\infty}+ \Big{\|}T_{m(n)}(y)-y\Big{\|}_{\infty}\\
 &\le  \Big{\|}y_n-y\Big{\|}_{\infty}+ \Big{\|}T_{m(n)}(y)-y\Big{\|}_{\infty},   
\end{align*}
where the inequality in the first term of the second line is due to the definition of $T_{m(n)}$ as in Lemma \ref{lem:3}. On the other hand for any $y\in \mathbb D_0=\mathcal C_u([0,1]^2)$, we have already established in Lemma \ref{lem:3} that, $\Big{\|}T_{m(n)}(y)-y\Big{\|}_{\infty}\to 0$ as $m(n)\to\infty$. Combining everything we have,
\[
\|y_n-y\|_{\infty}\to 0\;\text{with}\; y_n\in\mathbb{D}_n,\;y\in\mathbb{D}_0\;\implies \Big{\|}T_m(y_n)-y\Big{\|}_{\infty}\to 0\;\text{as}\; n\to\infty.
\]
Finally apply Lemma \ref{lem:5} to get,
\[
T_m\Big[\sqrt{n}\Big(\hat{C}_n(u,v)-C(u,v)\Big)\Big]\rightsquigarrow G_C\quad\text{in}\quad \ell^{\infty}([0,1]^2)\;\;\text{as}\;n\to\infty.
\]
Therefore we are done.\hfill $\square$

\section{Proofs of Main Theorems}\label{sec:ApA}
\subsection{Proof of Theorem \ref{thm:consistency_checkerboard}}\label{appAthm4.1}
 We would like to establish the following:
 \begin{align}\label{eqn:1}
&\sup_{(u,v)\in[0,1]^2}\Big{|}\hat{C}_n^{(m)}(u,v)-C(u,v)\Big{|}=o(1)\quad\text{almost surely},\;\text{as}\; n\to\infty.
 \end{align}
First note that
\begin{align}\label{eqn:2}
& \sup_{(u,v)\in[0,1]^2}\Big{|}\hat{C}_n^{(m)}(u,v)-C(u,v)\Big{|}\nonumber\\
&\le \sup_{(u,v)\in[0,1]^2}\Big{|}\hat{C}_n^{(m)}(u,v)-C^{(m)}(u,v)\Big{|}+\sup_{(u,v)\in[0,1]^2}\Big{|}C^{(m)}(u,v)-C(u,v)\Big{|}. \end{align}
Now recall the definition
\begin{align}\label{eqn:3}
&C^{(m)}(u,v)\nonumber\\
&=
\sum_{j=1}^{m}
\sum_{i=1}^{m}
\mathbf{1}_{\left(\frac{i-1}{m},\,\frac{i}{m}\right]}(u)\,
\mathbf{1}_{\left(\frac{j-1}{m},\,\frac{j}{m}\right]}(v)
\times
\Bigg[
(1-\mu(u))(1-\mu(v))
\,C\!\left(\tfrac{i-1}{m},\tfrac{j-1}{m}\right)\nonumber\\
&+(1-\mu(u))\,\mu(v)
\,C\!\left(\tfrac{i-1}{m},\tfrac{j}{m}\right)
+\mu(u)(1-\mu(v))
\,C\!\left(\tfrac{i}{m},\tfrac{j-1}{m}\right)
+\mu(u)\,\mu(v)
\,C\!\left(\tfrac{i}{m},\tfrac{j}{m}\right)
\Bigg],
\end{align}
with $(u,v)\in [0,1]^2$, $\mu(u)=m\Big[u-\frac{i-1}{m}\Big]$ and $\mu(v)=m\Big[v-\frac{j-1}{m}\Big]$. The empirical checkerboard copula $\hat{C}_n^{(m)}(u,v)$ is defined similarly by replacing the actual copula $C(\cdot,\cdot)$ with empirical version $\hat{C}_n(\cdot,\cdot)$. First we give an upper bound on the first term in \ref{eqn:2}. Fix some $i,j$ such that $\mathbf{1}_{\left(\frac{i-1}{m},\,\frac{i}{m}\right]}(u)\,
\mathbf{1}_{\left(\frac{j-1}{m},\,\frac{j}{m}\right]}(v)=1$. Therefore, simple algebra entail that
\begin{align}\label{eqn:4}
\hat{C}_n^{(m)}(u,v)-C^{(m)}(u,v)&=\sum_{a=0}^1\sum_{b=0}^{1}\Big\{1-a-(-1)^a\mu(u)\Big\}\Big\{1-b-(-1)^b\mu(v)\Big\}\nonumber\\
&\quad\times\Bigg[\hat{C}_n\Big(\frac{a+i-1}{m},\frac{b+j-1}{m}\Big)-C\Big(\frac{a+i-1}{m},\frac{b+j-1}{m}\Big)\Bigg],
\end{align}
where it is straightforward to see that
\[
\sum_{a=0}^1\sum_{b=0}^{1}\Big\{1-a-(-1)^a\mu(u)\Big\}\Big\{1-b-(-1)^b\mu(v)\Big\}=1.
\]
Let
\(
f(u,v)=\hat C_n(u,v)-C(u,v).\)
Then the checkerboard approximation satisfies
\[
\hat{C}_n^{(m)}(u,v)-C^{(m)}(u,v)
=
(T_m f)(u,v)
=
T_m\big(\hat C_n-C\big)(u,v),
\]
where define
\begin{align}
(T_m f)(u,v)
&:=
\sum_{a=0}^1\sum_{b=0}^1
\Big\{1-a-(-1)^a\mu(u)\Big\}
\Big\{1-b-(-1)^b\mu(v)\Big\}
f\!\left(\frac{a+i-1}{m},\frac{b+j-1}{m}\right).
\label{eq:Tm}
\end{align}
The operator $T_m$ is linear and norm--contractive, i.e., 
\begin{align}\label{eqn:6}
\|T_m f\|_\infty \le \|f\|_\infty=\Big{\|}\hat C_n(u,v)-C(u,v)\Big{\|}_\infty=O\Big[n^{-1/2}(\log\log n)^{1/2}\Big],\quad\text{almost surely},
\end{align}
where the last equality follows from Lemma 1 of \citet{janssen2012bernstein}. \\

Now we give an upper bound on the second term in \ref{eqn:1}. Note that it is possible to write
\[
C^{(m)}(u,v)-C(u,v)=T_m(C(u,v))-C(u,v)=[T_mC-C](u,v).
\]
Fix $(u,v)\in[0,1]^2$ and let
\(u_0=\frac{i-1}{m}, v_0=\frac{j-1}{m},\)
where $(u,v)\in\big(\tfrac{i-1}{m},\tfrac{i}{m}\big]\times
\big(\tfrac{j-1}{m},\tfrac{j}{m}\big]$.
Set
\[
h_u=u-u_0,\quad h_v=v-v_0,\quad u_1 = u_0 + \frac{1}{m},\quad \ v_1 = v_0 + \frac{1}{m}
\]
so that $|h_u|,|h_v|\le m^{-1}$. Using the fact that a copula $C$ is Lipschitz with respect to the $L^1$ norm, we have
\[
|C(u,v) - C(u',v')| \le |u-u'| + |v-v'|, \quad \forall (u,v),(u',v') \in [0,1]^2.
\]

In particular, for each vertex $(u_k,v_\ell)$ of the grid cell containing $(u,v)$, we obtain
\[
|C(u_k,v_\ell) - C(u,v)| \le |u_k - u| + |v_\ell - v| \le \frac{2}{m}.
\]

Since $T_m C(u,v)$ is a convex combination of the four values
$C(u_k,v_\ell)$, $k,\ell \in \{0,1\}$, it follows that
\[
\begin{aligned}
|T_m C(u,v) - C(u,v)|
&\le \sum_{k,\ell \in \{0,1\}} c_{k\ell}(u,v)\, |C(u_k,v_\ell) - C(u,v)| \\
&\le \sum_{k,\ell \in \{0,1\}} c_{k\ell}(u,v)\, \frac{2}{m}
= \frac{2}{m}.
\end{aligned}
\]

Therefore,
\begin{align}\label{eqn:7}
\|T_m C - C\|_\infty \le \frac{2}{m}.
\end{align}
Combining (\ref{eqn:6}) and (\ref{eqn:7}) we get,
\[
sup_{(u,v)\in[0,1]^2}\Big{|}\hat{C}_n^{(m)}(u,v)-C(u,v)\Big{|}=O\Bigg[\frac{(\log\log n)^{1/2}}{n^{1/2}}+\frac{1}{m}\Bigg]=o(1),\quad\text{almost surely},
\]
and hence the proof is complete. \hfill$\square$

\subsection{Proof of Theorem \ref{thm:weakconv_checkerboard}}\label{appAthm4.2}
Here we establish that,
\[
\sqrt{n}\Big[\hat{C}_n^{(m)}(u,v)-C(u,v)\Big]\rightsquigarrow G_C\quad\text{in}\quad \ell^{\infty}([0,1]^2)\;\;\text{as}\;n\to\infty.
\]
Recall the splitting as earlier,
\begin{align}\label{eqn:8}
\sqrt{n}\Big[\hat{C}_n^{(m)}(u,v)-C(u,v)\Big]&=\underbrace{\sqrt{n}\Big[\hat{C}_n^{(m)}(u,v)-C^{(m)}(u,v)\Big]}_{\text{stochastic term}}+\underbrace{\sqrt{n}\Big[C^{(m)}(u,v)-C(u,v)\Big]}_{\text{bias term}}\nonumber\\
&=T_m\Bigg[\sqrt{n}\Big[\hat{C}_n(u,v)-C(u,v)\Big]\Bigg]+\sqrt{n}\Big[C^{(m)}(u,v)-C(u,v)\Big]
\end{align}
From the (\ref{eqn:7}) of the proof of Theorem \ref{thm:consistency_checkerboard} we have
\[
\sup_{(u,v)\in[0,1]^2}\Big{|}\sqrt{n}\Big[C^{(m)}(u,v)-C(u,v)\Big]\Big{|}=O\Big[\frac{\sqrt{n}}{m}\Big]=o(1),
\]
since $\sqrt{n}=o(m)$. Now for the stochastic part, we apply Lemma \ref{lem:6} to conclude,
\[
T_m\Big[\sqrt{n}\Big(\hat{C}_n(u,v)-C(u,v)\Big)\Big]\rightsquigarrow G_C\quad\text{in}\quad \ell^{\infty}([0,1]^2)\;\;\text{as}\;n\to\infty.
\]
Therefore the proof is complete due to Slutsky's theorem. \hfill $\square$

\subsection{Proof of Theorem \ref{thm:strongconstail}}\label{appAthm4.5}
Write
\[
\hat \Lambda_{L,n}^{(m)}(x,y) - \Lambda_L(x,y) 
= \underbrace{\hat \Lambda_{L,n}^{(m)}(x,y) - \hat \Lambda_{L,n}(x,y)}_{\text{checkerboard approximation}}
+ \underbrace{\hat \Lambda_{L,n}(x,y) - \Lambda_L(x,y)}_{\text{empirical tail copula convergence}},
\]
where
\[
\hat \Lambda_{L,n}(x,y) = \frac{n}{k_n} \, \hat C_n\Big(\frac{k_n x}{n}, \frac{k_n y}{n}\Big)
\]
is the usual empirical lower tail copula. By Theorem 6 of \citet{schmidt2006}, we have
\[
\hat \Lambda_{L,n} \;\; \xrightarrow{\text{a.s.}} \;\; \Lambda_L \quad \text{in } B_\infty(\overline{\mathbb{R}}_+^2),\quad\text{i.e.},\quad
\mathbf{P} \Big[ \lim_{n\to\infty} d(\hat \Lambda_{L,n}, \Lambda_L) = 0 \Big] = 1,
\]
where the space $B_\infty(\overline{\mathbb{R}}_+^2)$ consists of all functions $f: \overline{\mathbb{R}}_+^2 \to \mathbb{R}$ that are locally uniformly bounded on every compact subset of $\overline{\mathbb{R}}_+^2$. It is equipped with the metric
\[
d(f_1,f_2) = \sum_{i=1}^{\infty} 2^{-i} \big( \| f_1 - f_2 \|_{T_i} \wedge 1 \big),
\]
where for all $i\in \mathbb N$,
\begin{align*}
T_{3i} &= T_{3i-1} \cup [0,i]^2,\quad T_{3i-1}= T_{3i-2} \cup ([0,i] \times \{\infty\}),\quad T_{3i-2}= T_{3(i-1)} \cup (\{\infty\} \times [0,i]), \\
T_0 &= \emptyset,\qquad\text{and}\qquad\;\|f_1 - f_2\|_{T_i} = \sup_{(x,y) \in T_i} |f_1(x,y)-f_2(x,y)|
\end{align*}

\medskip
Now, by definition,
\[
\Big(\widehat{\Lambda}_{L,n}^{(m)}(x,y) - \hat \Lambda_{L,n}(x,y)\Big)
=
\frac{n}{k_n}\,
\Big[
\hat C_n^{(m)}(u_n,v_n) - \hat C_n(u_n,v_n)
\Big],
\]
where $u_n = k_n x/n$, $v_n = k_n y/n$. Recall that definition of empirical checker board copula,
\begin{align}\label{eqn:43}
&\hat{C}_n^{(m)}(u,v)\nonumber\\
&=
\sum_{j=1}^{m}
\sum_{i=1}^{m}
\mathbf{1}_{\left(\frac{i-1}{m},\,\frac{i}{m}\right]}(u)\,
\mathbf{1}_{\left(\frac{j-1}{m},\,\frac{j}{m}\right]}(v)
\times
\Bigg[
(1-\mu(u))(1-\mu(v))
\,\hat{C}_n\!\left(\tfrac{i-1}{m},\tfrac{j-1}{m}\right)\nonumber\\
&+(1-\mu(u))\,\mu(v)
\,\hat{C}_n\!\left(\tfrac{i-1}{m},\tfrac{j}{m}\right)
+\mu(u)(1-\mu(v))
\,\hat{C}_n\!\left(\tfrac{i}{m},\tfrac{j-1}{m}\right)
+\mu(u)\,\mu(v)
\,\hat{C}_n\!\left(\tfrac{i}{m},\tfrac{j}{m}\right)
\Bigg],
\end{align}
with $(u,v)\in [0,1]^2$, $\mu(u)=m\Big[u-\frac{i-1}{m}\Big]$ and $\mu(v)=m\Big[v-\frac{j-1}{m}\Big]$. Let $(X_i,Y_i)_{i=1}^n$ be an i.i.d.\ sample from a bivariate distribution
with continuous marginal distribution functions.
Let $R_{Xi}$ and $R_{Yi}$ denote the ranks of $X_i$ and $Y_i$
among $X_1,\dots,X_n$ and $Y_1,\dots,Y_n$, respectively.
Define the pseudo--observations
\[
U_i=\frac{R_{Xi}}{n}, \qquad
V_i=\frac{R_{Yi}}{n}, \qquad i=1,\dots,n.
\]
The empirical copula $\hat{C}_n:[0,1]^2\to[0,1]$ is defined by
\[
\hat{C}_n(u,v)
=
\frac{1}{n}
\sum_{i=1}^n
\mathbf 1\!\left(
U_i\le u,\;
V_i\le v
\right),
\qquad (u,v)\in[0,1]^2.
\]
Fix $v\in[0,1]$ and assume without loss of generality that $u_1<u_2$.  
By definition,
\[
\hat C_n(u_2,v)-\hat C_n(u_1,v) = \frac{1}{n} \sum_{i=1}^n \mathbf 1(u_1 < U_i \le u_2,\, V_i \le v)\le \frac{1}{n} \sum_{i=1}^n \mathbf 1(u_1 < U_i \le u_2).
\]
Now assume $u_2-u_1\ge 1/n$. If $\lfloor nu \rfloor$ denotes the integer part of $nu$ for a particular $u$, then 
\[
\#\{i: u_1 < U_i \le u_2\} \le \lfloor n u_2 \rfloor - \lfloor n u_1 \rfloor \le n(u_2-u_1)+1,
\]
and hence dividing both side by $n$ we have
\[
\hat C_n(u_2,v)-\hat C_n(u_1,v) \le (u_2-u_1) + \frac{1}{n} \le 2(u_2-u_1).
\]
On the other hand, when $u_2-u_1 < 1/n$ then $\#\{i: u_1 < U_i \le u_2\} \le 1$. The arguments for the  other coordinate is analogous. Thus we have
\begin{align}\label{eqn:100}
|\hat C_n(u_2,v_2)-\hat C_n(u_1,v_1)| &\le |\hat C_n(u_2,v_2)-\hat C_n(u_1,v_2)| + |\hat C_n(u_1,v_2)-\hat C_n(u_1,v_1)|\nonumber \\
&\le 2(|u_2-u_1| + |v_2-v_1|) + 2/n.
\end{align}
Fix $(u,v)\in[0,1]^2$. Then there exist $i,j \in \{1, \dots, m\}$ such that
\[
u \in \Big(\frac{i-1}{m}, \frac{i}{m}\Big], \quad
v \in \Big(\frac{j-1}{m}, \frac{j}{m}\Big].
\]
Then from \eqref{eqn:43} we can write
\begin{align}\label{eq:convex_comb}
\hat C_n^{(m)}(u,v) - \hat C_n(u,v)
= \sum_{a=0}^{1} \sum_{b=0}^{1} \alpha_{ab}(u,v) \Bigg[
\hat C_n\Big(\frac{i-1+a}{m},\frac{j-1+b}{m}\Big) - \hat C_n(u,v)
\Bigg],
\end{align}
where
\[
\alpha_{ab}(u,v) = (1-a-(-1)^a \mu(u)) (1-b-(-1)^b \mu(v)) \ge 0, \quad 
\sum_{a=0}^{1} \sum_{b=0}^{1} \alpha_{ab}(u,v) = 1.
\]
Moreover, due to (\ref{eqn:100}) we have 
\begin{align}\label{eqn:kjnj}
|\hat C_n(u',v') - \hat C_n(u,v)| \le 4/m + 2/n.
\end{align}
Note that, this bound holds path wise for any fixed realization $\omega$. Rescaling to the tail copula gives
\[
\sup_{(x,y)\in[0,\infty)^2} \big|\hat \Lambda_{L,n}^{(m)}(x,y) - \hat \Lambda_{L,n}(x,y)\big| \le \frac{4n}{k_n m}+\frac{2}{k_n}.
\]
The condition $\frac{n}{\sqrt{k_n}}=o(m)$ ensures that this term vanishes as $n \to \infty$. Now, convergence of elements of $B_\infty(\overline{\mathbb{R}}_+^2)$ in that metric is equivalent to uniform convergence on each compact $T_i$. Using the triangle inequality for the metric $d$:
\[
d(\hat \Lambda_{L,n}^{(m)}, \Lambda_L) \le d(\hat \Lambda_{L,n}^{(m)}, \hat \Lambda_{L,n}) + d(\hat \Lambda_{L,n}, \Lambda_L).
\]

Hence,
\[
d(\hat \Lambda_{L,n}^{(m)}, \Lambda_L) \to 0 \quad \text{almost surely}.
\]
Therefore we are done.\hfill $\square$

\subsection{Proof of Theorem \ref{thm:weakconv_tail_checkerboard}}\label{appAthm4.3}
We only establish the weak convergence of the lower tail process $\{\hat \Lambda_{L,n}(x,y)\}_{n\geq 1}$. The proof for the upper tail one is similar.
Recall that,
\[
\Lambda_L(x,y)
=
\lim_{t \downarrow 0}
\frac{C(tx,\, ty)}{t},
\qquad x,y \ge 0 .
\]
Define the empirical tail copula
\[
\hat \Lambda_{L,n}(x,y)
:=
\frac{n}{k_n}\,
\hat C_n\!\left(\frac{k_n x}{n},\frac{k_n y}{n}\right),
\qquad x,y\ge 0.
\]
The estimator based on empirical checkerboard copula is given by,
\[
\widehat{\Lambda}_{L,n}^{(m)}(x,y)
=
\frac{n}{k_n}
\hat C_n^{(m)}\!\left(\frac{k_n x}{n},\frac{k_n y}{n}\right),
\qquad x,y \ge 0 .
\]
Now we can write,
 \begin{equation}\label{eq:checkerboard-decomp}
\sqrt{k_n} \Big(
\widehat{\Lambda}_{L,n}^{(m)}(x,y) - \Lambda_L(x,y)
\Big)
=
\underbrace{\sqrt{k_n} \Big(
\hat \Lambda_{L,n}(x,y) - \Lambda_L(x,y)
\Big)}_{\text{empirical process term}}
+
\underbrace{\sqrt{k_n} \Big(
\widehat{\Lambda}_{L,n}^{(m)}(x,y) - \hat \Lambda_{L,n}(x,y)
\Big)}_{\text{checkerboard approximation}}.
\end{equation}
Under the regularity condition (C.1) and (C.2), Theorem 2.2 of \citet{BucherDette2013} entails
\[
\sqrt{k_n} \Big(
\hat \Lambda_{L,n}(x,y) - \Lambda_L(x,y)
\Big)\rightsquigarrow \mathbb{G}_{\hat{\Lambda}_L}(x,y).
\]
Therefore it is enough to show that, 
\begin{align}\label{eqn:101}
\sup_{(x,y)\in[0,\infty)^2}\sqrt{k_n}\,\Big|\widehat{\Lambda}_{L,n}^{(m)}(x,y) - \hat \Lambda_{L,n}(x,y)\Big| \le \frac{4n}{m\sqrt{k_n}}+\frac{2}{\sqrt{k_n}},
\end{align}
since $\frac{n}{\sqrt{k_n}}=o(m)$. Now, by definition,
\[
\sqrt{k_n}\,\Big(\widehat{\Lambda}_{L,n}^{(m)}(x,y) - \hat \Lambda_{L,n}(x,y)\Big)
=
\frac{n}{\sqrt{k_n}}\,
\Big[
\hat C_n^{(m)}(u_n,v_n) - \hat C_n(u_n,v_n)
\Big],
\]
Now we follow the steps of Theorem \ref{thm:strongconstail} till equation (\ref{eqn:kjnj}) to obtain the desired (\ref{eqn:101}). 
The proof is complete applying Slutsky's theorem. \hfill $\square$

\subsection{Proof of Theorem \ref{thm:bootcopula}}\label{sec:klj}
Note the following decomposition: 
\begin{align*}
&\frac{\mu}{\tau}\sqrt{n}\Big(\hat C_n^{(m),\xi,\xi}-\hat{C}_n^{(m)}\Big)\\
&=\frac{\mu}{\tau}\sqrt{n}\Big(\hat C_n^{\xi,\xi}-\hat{C}_n\Big)+\frac{\mu}{\tau}\sqrt{n}\Big(\hat C_n^{(m),\xi,\xi}-\hat{C}_n^{\xi,\xi}\Big)-\frac{\mu}{\tau}\sqrt{n}\Big(\hat C_n^{(m)}-\hat{C}_n\Big)
\end{align*}
By Theorem 2.4 of \citet{Bucher2011}, it holds conditionally on the data that,
\[
\frac{\mu}{\tau}\sqrt{n}\Big(\hat C_n^{\xi,\xi}-\hat{C}_n\Big)\overset{\mathbf P}{\underset{\xi}{\rightsquigarrow}}\;\ 
\mathbb G_{C}
\quad
\text{in } \ell^\infty([0,1]^2)
\]
Now by equation (\ref{eqn:kjnj}), we can conclude that,
\begin{align*}
 \frac{\mu}{\tau}\sqrt{n}\sup_{(u,v)\in[0,1]^2} \Big|\hat C_n^{(m)}(u,v)-\hat{C}_n(u,v)\Big|=O_p\Big[\frac{\sqrt{n}}{m}+\frac{1}{\sqrt{n}}\Big]=o_p(1), 
\end{align*}
due the assumption $\sqrt{n}=o(m)$. Next note that,
\begin{align*}
&\sup_{(u,v)\in[0,1]^2}\Big{|}\widehat{C}_n^{(m),\xi,\xi}(u,v)-\widehat{C}_n^{\xi,\xi}(u,v)\Big{|}\\
&=\sup_{(u,v)\in[0,1]^2}\Bigg{|}\sum_{a=0}^{1} \sum_{b=0}^{1} \alpha_{ab}(u,v) \Bigg[
\widehat{C}_n^{\xi,\xi}\Big(\frac{i-1+a}{m},\frac{j-1+b}{m}\Big) - \widehat{C}_n^{\xi,\xi}(u,v)
\Bigg]\Bigg{|},
\end{align*}
where
\[
\alpha_{ab}(u,v) = (1-a-(-1)^a \mu(u)) (1-b-(-1)^b \mu(v)) \ge 0, \quad 
\sum_{a=0}^{1} \sum_{b=0}^{1} \alpha_{ab}(u,v) = 1.
\]
Fix \(v \in [0,1]\). Let \(0 \le u_1 < u_2 \le 1\).
By definition of the left–continuous generalized inverse of a distribution
function, the map \(p \mapsto (F_n^\xi)^{-}(p)\) is non-decreasing. Hence,
\[
(F_n^\xi)^{-}(u_1) \le (F_n^\xi)^{-}(u_2).
\]
Therefore, for every \(i=1,\dots,n\),
\[
\mathbf 1\!\left\{
X_i \le (F_n^\xi)^{-}(u_1),\;
Y_i \le (G_n^\xi)^{-}(v)
\right\}
\;\le\;
\mathbf 1\!\left\{
X_i \le (F_n^\xi)^{-}(u_2),\;
Y_i \le (G_n^\xi)^{-}(v)
\right\}.
\]

Multiplying both sides by the positive weight \(\xi_i/\bar{\xi}_n\) and
summing over \(i=1,\dots,n\), we obtain
\begin{align*}
&\frac{1}{n}
\sum_{i=1}^n
\frac{\xi_i}{\bar{\xi}_n}
\mathbf 1\!\left\{
X_i\le (F_n^\xi)^{-}(u_1),\;
Y_i \le (G_n^\xi)^{-}(v)
\right\}\\
&\qquad\qquad\;\le\;
\frac{1}{n}
\sum_{i=1}^n
\frac{\xi_i}{\bar{\xi}_n}
\mathbf 1\!\left\{
X_i \le (F_n^\xi)^{-}(u_2),\;
Y_i \le (G_n^\xi)^{-}(v)
\right\}.
\end{align*}
That is,
\(\widehat{C}_n^{\xi,\xi}(u_1,v)
\le
\widehat{C}_n^{\xi,\xi}(u_2,v),
\) which proves that \(u \mapsto \widehat{C}_n^{\xi,\xi}(u,v)\) is nondecreasing
for every fixed \(v \in [0,1]\). Conclusion with respect to coordinate $v$ for fixed $u$ is similar. This proves monotonicity.

\medskip

Fix \(v \in [0,1]\) and let \(0 \le u_1 < u_2 \le 1\).
By definition,
\begin{align*}
\widehat{C}_n^{\xi,\xi}(u_2,v)
-
\widehat{C}_n^{\xi,\xi}(u_1,v)
&=
\frac{1}{n}
\sum_{i=1}^n
\frac{\xi_i}{\bar{\xi}_n}
\Bigl[
\mathbf 1\!\left\{
X_i \le (F_n^\xi)^{-}(u_2),\;
Y_i \le (G_n^\xi)^{-}(v)
\right\}
\\
&\hspace{4em}
-
\mathbf 1\!\left\{
X_i \le (F_n^\xi)^{-}(u_1),\;
Y_i \le (G_n^\xi)^{-}(v)
\right\}
\Bigr].
\end{align*}

Since \((F_n^\xi)^{-}\) is nondecreasing, the difference of indicators can be
nonzero only if
\[
(F_n^\xi)^{-}(u_1) < X_i \le (F_n^\xi)^{-}(u_2)
\quad\text{and}\quad
Y_i \le (G_n^\xi)^{-}(v).
\]
Hence,
\begin{align*}
0& \le
\mathbf 1\!\left\{
X_i \le (F_n^\xi)^{-}(u_2),\;
Y_i \le (G_n^\xi)^{-}(v)
\right\}
-
\mathbf 1\!\left\{
X_i \le (F_n^\xi)^{-}(u_1),\;
Y_i \le (G_n^\xi)^{-}(v)
\right\}\nonumber\\
&\qquad\qquad\qquad\qquad\le
\mathbf 1\!\left\{
(F_n^\xi)^{-}(u_1) < X_i \le (F_n^\xi)^{-}(u_2)
\right\}.    
\end{align*}

Therefore,
\begin{align*}
\widehat{C}_n^{\xi,\xi}(u_2,v)
-
\widehat{C}_n^{\xi,\xi}(u_1,v)
&\le
\frac{1}{n}
\sum_{i=1}^n
\frac{\xi_i}{\bar{\xi}_n}
\mathbf 1\!\left\{
(F_n^\xi)^{-}(u_1) < X_i \le (F_n^\xi)^{-}(u_2)
\right\}
\\
&=
F_n^\xi\!\bigl((F_n^\xi)^{-}(u_2)\bigr)
-
F_n^\xi\!\bigl((F_n^\xi)^{-}(u_1)\bigr).
\end{align*}

Set $x_u := (F_n^\xi)^{-}(u)$. By definition of the generalized inverse,
\[
F_n^\xi(x_u-) < u \le F_n^\xi(x_u),
\]
where $F_n^\xi(x_u-)$ denotes the left limit at $x_u$. Since $F_n^\xi$ is a step function with jumps of size at most \(
\Delta_n := \max_{1 \le i \le n} \frac{\xi_i}{n\bar{\xi}_n}.
\) we can write
\[
F_n^\xi(x_u) = F_n^\xi(x_u-) + J(x_u),
\]
where $J(x_u)$ is the jump size at $x_u$, satisfying $0 \le J(x_u) \le \Delta_n$. Combining these observations yields
\[
u \le F_n^\xi\big((F_n^\xi)^{-}(u)\big)
= F_n^\xi(x_u)
\le F_n^\xi(x_u-) + \Delta_n
< u + \Delta_n.
\]
Therefore, for all $u \in [0,1]$,
\(u \le F_n^\xi\big((F_n^\xi)^{-}(u)\big) \le u + \Delta_n.
\) Consequently,
\begin{align}\label{eqn:lkjh}
\widehat{C}_n^{\xi,\xi}(u_2,v)
-
\widehat{C}_n^{\xi,\xi}(u_1,v)
\le
(u_2 - u_1) + \Delta_n.
\end{align}
By construction, \((\xi_i)_{i\ge1}\) be i.i.d.\ positive sub-exponential random variables with mean
\(0<\mu<\infty\). Then it's easy to verify that, $\max_{1\le i\le n}\xi_i = O_p(\log n)$ (cf. Lemma 2.2.10 of \citet{vdVWellner1996}).
By the weak law of large numbers,
\(\bar{\xi}_n \xrightarrow{\mathbf P} \mu\) and hence
\(
\bar{\xi}_n = O_p(1).
\)
Combining these we get; $\Delta_n=O_p\Big(\frac{\log n}{n}\Big)$. Similar argument holds for $v$ coordinate with fixed $u$. For any $(u,v)\in[0,1]^2$, let
\[
u\in\left(\frac{i-1}{m},\frac{i}{m}\right],\qquad
v\in\left(\frac{j-1}{m},\frac{j}{m}\right].
\]
Using the monotonicity of $\widehat{C}_n^{\xi,\xi}$ in each coordinate together with
\eqref{eqn:lkjh}, we obtain
\begin{align*}
0
&\le
\widehat{C}_n^{\xi,\xi}(u,v)
-
\widehat{C}_n^{\xi,\xi}\!\left(\frac{i-1}{m},\frac{j-1}{m}\right)\\
&=
\Bigg[
\widehat{C}_n^{\xi,\xi}(u,v)
-
\widehat{C}_n^{\xi,\xi}\!\left(\frac{i-1}{m},v\right)
\Bigg]
+
\Bigg[
\widehat{C}_n^{\xi,\xi}\!\left(\frac{i-1}{m},v\right)
-
\widehat{C}_n^{\xi,\xi}\!\left(\frac{i-1}{m},\frac{j-1}{m}\right)
\Bigg]\\
&\le
\left(u-\frac{i-1}{m}\right)+\Delta_n
+
\left(v-\frac{j-1}{m}\right)+\Delta_n\\
&\le
\frac{2}{m}+2\Delta_n.
\end{align*}
Similarly,
\begin{align*}
0
&\le
\widehat{C}_n^{\xi,\xi}\!\left(\frac{i}{m},\frac{j}{m}\right)
-
\widehat{C}_n^{\xi,\xi}(u,v)\\
&=
\Bigg[
\widehat{C}_n^{\xi,\xi}\!\left(\frac{i}{m},\frac{j}{m}\right)
-
\widehat{C}_n^{\xi,\xi}\!\left(u,\frac{j}{m}\right)
\Bigg]
+
\Bigg[
\widehat{C}_n^{\xi,\xi}\!\left(u,\frac{j}{m}\right)
-
\widehat{C}_n^{\xi,\xi}(u,v)
\Bigg]\\
&\le
\left(\frac{i}{m}-u\right)+\Delta_n
+
\left(\frac{j}{m}-v\right)+\Delta_n\\
&\le
\frac{2}{m}+2\Delta_n.
\end{align*}
The remaining two corner points,
\(\left(\frac{i-1}{m},\frac{j}{m}\right)
\;\text{and}\;
\left(\frac{i}{m},\frac{j-1}{m}\right),
\) are handled analogously, yielding
\[
\max_{a,b\in\{0,1\}}
\left|
\widehat{C}_n^{\xi,\xi}\!\left(\frac{i-1+a}{m},\frac{j-1+b}{m}\right)
-
\widehat{C}_n^{\xi,\xi}(u,v)
\right|
\le
\frac{2}{m}+2\Delta_n.
\]
Therefore,
\begin{align*}
&\left|
\widehat{C}_n^{(m),\xi,\xi}(u,v)
-
\widehat{C}_n^{\xi,\xi}(u,v)
\right|\\
&\le
\sum_{a=0}^{1}\sum_{b=0}^{1}
\alpha_{ab}(u,v)
\left|
\widehat{C}_n^{\xi,\xi}\!\left(\frac{i-1+a}{m},\frac{j-1+b}{m}\right)
-
\widehat{C}_n^{\xi,\xi}(u,v)
\right|\\
&\le
\left(\sum_{a=0}^{1}\sum_{b=0}^{1}\alpha_{ab}(u,v)\right)
\left(\frac{2}{m}+2\Delta_n\right)
=
\frac{2}{m}+2\Delta_n.
\end{align*}
Since the above bound is uniform over $(u,v)\in[0,1]^2$, it follows that
\begin{align}\label{eqn:asdf}
\sup_{(u,v)\in[0,1]^2}
\left|
\widehat{C}_n^{(m),\xi,\xi}(u,v)
-
\widehat{C}_n^{\xi,\xi}(u,v)
\right|
\le
\frac{2}{m}+2\Delta_n.
\end{align}
Recalling that
\(\Delta_n
=
O_p\!\left(\frac{\log n}{n}\right),
\) we conclude that
\[
\frac{\mu}{\tau}\sqrt{n}\sup_{(u,v)\in[0,1]^2}
\left|
\widehat{C}_n^{(m),\xi,\xi}(u,v)
-
\widehat{C}_n^{\xi,\xi}(u,v)
\right|
=
O_p\!\left(\frac{\sqrt{n}}{m}+\frac{\log n}{\sqrt{n}}\right)=o_p(1),
\]
provided $\sqrt{n}=o(m)$. Therefore the proof is complete due to Slutsky's theorem. \hfill $\square$
\subsection{Proof of Theorem \ref{thm:cb-dm-bootstrap}}\label{appAthm4.6}
 We decompose
\begin{align*}
\alpha_n^{(m)}
&=
\frac{\mu}{\tau}\sqrt{k_n}
\Bigl(
\widehat{\Lambda}^{(m),\xi,\xi}_{L,n}
-
\widehat{\Lambda}^{(m)}_{L,n}
\Bigr)
\\[0.4em]
&=
\frac{\mu}{\tau}\sqrt{k_n}
\Bigl(
\widehat{\Lambda}^{\xi,\xi}_{L,n}
-
\widehat{\Lambda}_{L,n}
\Bigr)
+
\frac{\mu}{\tau}\sqrt{k_n}
\Bigl(
\widehat{\Lambda}^{(m),\xi,\xi}_{L,n}
-
\widehat{\Lambda}^{\xi,\xi}_{L,n}
\Bigr)
-
\frac{\mu}{\tau}\sqrt{k_n}
\Bigl(
\widehat{\Lambda}^{(m)}_{L,n}
-
\widehat{\Lambda}_{L,n}
\Bigr).
\end{align*}
By Theorem~3.4 of \citet{BucherDette2013}, it holds conditionally on the data that
\[
\frac{\mu}{\tau}\sqrt{k_n}
\Bigl(
\widehat{\Lambda}^{\xi,\xi}_{L,n}
-
\widehat{\Lambda}_{L,n}
\Bigr)
\;\overset{\mathbf P}{\underset{\xi}{\rightsquigarrow}}\;\ 
\mathbb G_{\hat\Lambda_L}
\quad
\text{in } B_\infty(\overline{\mathbb R}_+^2).
\]
An exact similar calculation as in the proof of Theorem \ref{thm:weakconv_tail_checkerboard}, will give us that,
\begin{align*}
\frac{\mu}{\tau}\sqrt{k_n}
\sup_{(x,y)\in K}\Big{|}
\widehat{\Lambda}^{(m)}_{L,n}
-
\widehat{\Lambda}_{L,n}\Big{|}=O_p\Bigg[\frac{n}{m\sqrt{k_n}}+\frac{1}{\sqrt{k_n}}\Bigg]=o_p(1),
\end{align*}
provided $\frac{n}{\sqrt{k_n}}=o(m)$. All it remains to handle the middle term. Now fix some $(x,y)\in K$. Based on the decomposition of unconditional third term, we will write;
\begin{align*}
&\frac{\mu}{\tau}\sqrt{k_n}
\Bigl(
\widehat{\Lambda}^{(m),\xi,\xi}_{L,n}(x,y)
-
\widehat{\Lambda}^{\xi,\xi}_{L,n}(x,y)
\Bigr)\\
&=\frac{\mu}{\tau}\frac{n}{\sqrt{k_n}}\Big[\widehat{C}_n^{(m),\xi,\xi}(u_n,v_n)-\widehat{C}_n^{\xi,\xi}(u_n,v_n)\Big],\qquad\text{where}\; u_n=\frac{k_nx}{n},\;v_n=\frac{k_ny}{n}\\
&\le \frac{\mu}{\tau}\frac{n}{\sqrt{k_n}}\sup_{(u,v)\in[0,1]^2}\Big{|}\widehat{C}_n^{(m),\xi,\xi}(u,v)-\widehat{C}_n^{\xi,\xi}(u,v)\Big{|}
\end{align*}
Following equation (\ref{eqn:asdf}) and tracing back to original requirement we get,
\begin{align*}
&\frac{\mu}{\tau}\sqrt{k_n}\sup_{(x,y)\in K}
\Big|
\widehat{\Lambda}^{(m),\xi,\xi}_{L,n}(x,y)
-
\widehat{\Lambda}^{\xi,\xi}_{L,n}(x,y)
\Big|=O_p\Big[\frac{n}{m\sqrt{k_n}}+\frac{\log n}{\sqrt{k_n}}\Big]=o_p(1),
\end{align*}
since $\frac{n}{\sqrt{k_n}}=o(m)$ and $(\log n)^2=o(k_n)$ hold. Finally apply Slutsky's theorem to conclude the proof. \hfill $\square$
\subsection{Proof of Theorem \ref{thm:test_checkerboard}}\label{appAthm6.1}
First note that due to Theorem~\ref{thm:weakconv_tail_checkerboard}, the independence of the samples $\{(X_i, Y_i)\}_{i=1}^{n_1}$ and $\{(X_i^\prime, Y_i^\prime)\}_{i=1}^{n_2}$ and the Slutsky's theorem, 
\begin{align*}
\mathcal E_n
&=
\sqrt{\frac{k_{n,2}}
{k_{n,1}+k_{n,2}}}
\sqrt{k_{n,1}}
\left(
\widehat{\Lambda}^{(m_1)}_{L,n_1}
-
\Lambda_L
\right)
-
\sqrt{\frac{k_{n,1}}
{k_{n,1}+k_{n,2}}}
\sqrt{k_{n,2}}
\left(
\widehat{\Lambda}^{\prime(m_2)}_{L,n_2}
-
\Lambda_L^\prime
\right)\nonumber\\
&\rightsquigarrow \mathcal E
:=
\sqrt{1-\zeta}\,
\mathbb G_{\hat\Lambda_L}
-
\sqrt{\zeta}\,
\mathbb G_{\hat\Lambda_L'}'\;\text{in}\;\mathcal B_\infty(\overline{\mathbb R}_+^2),
\end{align*}
under both $H_0$ and $H_1$. Hence by continuous mapping theorem (cf. Theorem 1.3.6 of \citet{vdVWellner1996}) we have
\begin{align}\label{eqn:6.10}
    \int_{0}^{\pi/2}\mathcal E_n^2(x_\phi)d\phi \rightsquigarrow \int_{0}^{\pi/2}\mathcal E^2(x_\phi)d\phi,
\end{align}
under both $H_0$ and $H_1$, where $x_\phi = (cos \phi, sin \phi)^\prime$. Similarly, we have
\begin{align}\label{eqn:6.11}
    \rho_n^*(\Lambda_L, \Lambda_L^\prime) \overset{\mathbf P}{\underset{\xi, \zeta} {\rightsquigarrow}} \int_{0}^{\pi/2}\mathcal E^2(x_\phi)d\phi,
\end{align}
irrespective of $H_0$ or $H_1$.\\ 

Now let us focus on part (a). Note that under $H_0$, $ \hat{\rho}_n(\Lambda_L, \Lambda_L^\prime) =\int_{0}^{\pi/2}\mathcal E_n^2(x_\phi)d\phi$ and hence from (\ref{eqn:6.10}) and (\ref{eqn:6.11}) we have
\begin{align}\label{eqn:6.12}
    \sup_{x \geq 0}\Big|\mathbf P_{\xi, \zeta}\Big(\rho_n^*(\Lambda_L, \Lambda_L^\prime) \leq x\Big) - \mathbf P\Big(\hat{\rho}_n(\Lambda_L, \Lambda_L^\prime) \leq x\Big)\Big| \xrightarrow{\mathbf P} 0, \; \text{as}\; n \rightarrow \infty,
\end{align}
under $H_0$, due to Polya's theorem, where $\mathbf P_{\xi, \zeta}(\cdot)$ denotes the conditional probability of the multiplier $(\xi_1, \zeta_1)$ given the observed samples. Therefore, under $H_0$,
\begin{align*}
   \Big|\mathbf E\psi_{n, \alpha}(\Lambda_L, \Lambda_L^\prime) -\alpha\Big| &= \Big|\mathbf P\big(\hat{\rho}_n(\Lambda_L, \Lambda_L^\prime) > \hat{q}_{n, 1-\alpha}\big) -\alpha\Big|\nonumber\\
   & \leq \mathbf E\bigg[\sup_{x \geq 0}\Big|\mathbf P_{\xi, \zeta}\Big(\rho_n^*(\Lambda_L, \Lambda_L^\prime) \leq x\Big) - \mathbf P\Big(\hat{\rho}_n(\Lambda_L, \Lambda_L^\prime) \leq x\Big)\Big|\bigg]\nonumber\\
   & \rightarrow 0,\; \text{as} \;n \rightarrow \infty. 
\end{align*}
The convergence in the last line is true due to (\ref{eqn:6.12}) and the dominated convergence theorem. Hence the part (a) follows. 
Now let us concentrate on part (b). First note that due to Theorem \ref{thm:strongconstail} and continuous mapping theorem we have
\begin{align*}
   &\int_{0}^{\pi/2}\big(\hat{\Lambda}_{L, n_1}^{(m_1)}(x_\phi) - \hat{\Lambda}_{L, n_2}^{\prime(m_2)}(x_\phi)\big)^2d\phi \xrightarrow{\mathbf P}   \int_{0}^{\pi/2}\big(\Lambda_L(x_\phi) - \hat{\Lambda}_{L}^\prime(x_\phi)\big)^2d\phi > 0\nonumber\\
   \Rightarrow & \int_{0}^{\pi/2}\mathcal E_n^2(x_\phi)d\phi \xrightarrow{\mathbf P} \infty,
\end{align*}
under $H_1$, since $k_{n, 1}k_{n, 2}/(k_{n, 1}+ k_{n, 2})\rightarrow \infty$. On the other hand, (\ref{eqn:6.11}) implies that $\hat q_{n, 1-\alpha}= O_P(1)$, as $n\rightarrow \infty$. Thus under $H_1$ we have 
\begin{align*}
   \mathbf E\psi_{n, \alpha}(\Lambda_L, \Lambda_L^\prime) &= \mathbf P\big(\hat{\rho}_n(\Lambda_L, \Lambda_L^\prime) > \hat{q}_{n, 1-\alpha}\big)  \rightarrow 1,\; \text{as} \;n \rightarrow \infty, 
\end{align*}
and hence the proof of part (b) is complete. $\hfill \square$

\subsection{Proof of Theorem \ref{thm:MD}}\label{sec:kjhg}
By Theorem~\ref{thm:strongconstail},
\[
d\!\left(\widehat{\Lambda}_{L,n}^{(m)},\Lambda_L\right)
\xrightarrow{\mathrm{a.s.}}0.
\]
Since the metric $d$ induces the topology of
$B_\infty(\overline{\mathbb{R}}_+^2)$, it follows that
\[
\sup_{\phi\in[0,\pi/2]}
\left|
\widehat{\Lambda}_{L,n}^{(m),\angle}(\phi)
-
\Lambda_L^{\angle}(\phi)
\right|
\xrightarrow{\mathrm{a.s.}}0.
\]
Denote, \(Q_n(\theta)
=
\int_0^{\pi/2}
\left(
\widehat{\Lambda}_{L,n}^{(m),\angle}(\phi)
-
\Lambda_L^{\angle}(\phi;\theta)
\right)^2
\,d\phi.
\) Hence, $\psi_n(\theta):=\partial_\theta Q_n(\theta)
=
-2
\int_0^{\pi/2}
\delta_\theta^{\angle}(\phi)
\left(
\widehat{\Lambda}_{L,n}^{(m),\angle}(\phi)
-
\Lambda_L^{\angle}(\phi;\theta)
\right)
\,d\phi.$ Similarly define,
\[\psi(\theta)
=
-2
\int_0^{\pi/2}
\delta_\theta^{\angle}(\phi)
\left(
\Lambda_L^{\angle}(\phi)
-
\Lambda_L^{\angle}(\phi;\theta)
\right)
\,d\phi.
\] Therefore,
\[
\psi_n(\theta)-\psi(\theta)
=
-2
\int_0^{\pi/2}
\delta_\theta^{\angle}(\phi)
\left(
\widehat{\Lambda}_{L,n}^{(m),\angle}(\phi)
-
\Lambda_L^{\angle}(\phi)
\right)
\,d\phi,
\]
which implies
\[
\begin{aligned}
\sup_{\theta\in\Theta}
\|\psi_n(\theta)-\psi(\theta)\|
&\le
2
\sup_{\theta\in\Theta}
\int_0^{\pi/2}
\|\delta_\theta^{\angle}(\phi)\|
\left|
\widehat{\Lambda}_{L,n}^{(m),\angle}(\phi)
-
\Lambda_L^{\angle}(\phi)
\right|
\,d\phi\\
&\le
2
\int_0^{\pi/2}
\sup_{\theta\in\Theta}
\|\delta_\theta^{\angle}(\phi)\|
\left|
\widehat{\Lambda}_{L,n}^{(m),\angle}(\phi)
-
\Lambda_L^{\angle}(\phi)
\right|
\,d\phi.
\end{aligned}
\] Consequently, by assumption (A.1),
\[
\begin{aligned}
\sup_{\theta\in\Theta}
\|\psi_n(\theta)-\psi(\theta)\|
&\le
2
\sup_{\phi\in[0,\pi/2]}
\left|
\widehat{\Lambda}_{L,n}^{(m),\angle}(\phi)
-
\Lambda_L^{\angle}(\phi)
\right|
\int_0^{\pi/2}
\sup_{\theta\in\Theta}
\|\delta_\theta^{\angle}(\phi)\|
\,d\phi
\xrightarrow{\mathrm{a.s.}}0.
\end{aligned}
\]
Hence due to this fact and assumption (A.2),  Theorem~5.9 of \citet{vanDerVaart2000} implies that
\[
\widehat{\theta}_n^{MD}
\xrightarrow{\mathbf P}
\theta_0.
\]
Next, the Taylor's theorem on $\psi_n(\theta)$ around $\theta_0$ gives
\[
0
=
\psi_n\!\left(\widehat{\theta}_n^{MD}\right)
=
\psi_n(\theta_0)
+
\partial_\theta\psi_n(\bar{\theta}_n)
\left(
\widehat{\theta}_n^{MD}
-
\theta_0
\right),
\]
where
\(\|\bar{\theta}_n-\theta_0\|
\le
\|\widehat{\theta}_n^{MD}-\theta_0\|.
\) Using the consistency of $\widehat{\theta}_n^{MD}$, the continuity assumptions (A.1)--(A.4), the definition of $A_{\theta_0}$ and Theorem~\ref{thm:weakconv_tail_checkerboard}, one can follow the arguments of the proof of Theorem 4.7 of \citet{Bucher2011}, to conclude
\[
\hat{\Theta}_n^{MD}
=
\sqrt{k_n}
\bigl(
\widehat{\theta}_n^{MD}
-
\theta_0
\bigr)
\rightsquigarrow
\Theta^{MD}.
\]
The formula of the covariance of the limiting process follows directly from the covariance structure of the Gaussian process $\mathbb{G}_{\widehat{\Lambda}_L}$, while the weak convergence of the bootstrap version $\Theta^{*MD}_n$ follows through the same line of arguments, but with the help of Theorem~\ref{thm:cb-dm-bootstrap} in place of Theorem~\ref{thm:weakconv_tail_checkerboard}. This completes the proof.\hfill $\square$

\end{appendix}

\bibliographystyle{amsplain}

\end{document}